\documentclass[12pt]{article}
\usepackage{amsmath}
\usepackage{amssymb}
\usepackage{graphicx}
\usepackage{bm}
\usepackage{algorithm}
\usepackage{algpseudocode}
\usepackage{natbib}
\usepackage{booktabs,siunitx,makecell,threeparttable}
\usepackage[font=footnotesize,labelfont=bf]{caption}
\usepackage{multirow}
\usepackage{multicol}
\usepackage{url} 
\newtheorem{assumption}{Assumption}
\newcommand{\blind}{0}
\newcommand{\lcbk}{\left\{}
\newcommand{\rcbk}{\right\}}
\newcommand{\lsbk}{\left[}
\newcommand{\rsbk}{\right]}
\newcommand{\lpth}{\left(}
\newcommand{\rpth}{\right)}

\newcommand{\midsepremove}{\aboverulesep = 0mm \belowrulesep = 0mm}
\newcommand{\midsepdefault}{\aboverulesep = 0.605mm \belowrulesep = 0.984mm}

\newcommand{\bX}{ \mbox{\bf X}}

\newcommand{\bW}{ \mbox{\bf W}}

\newcommand{\beq}{ \begin{equation}}
\newcommand{\eeq}{ \end{equation}}
\newcommand{\beqn}{ \begin{eqnarray}}
\newcommand{\eeqn}{ \end{eqnarray}}

\begin{document}

\def\spacingset#1{\renewcommand{\baselinestretch}%
{#1}\small\normalsize} \spacingset{1}


\if0\blind
{
  \title{\bf A Bayesian Semiparametric Method For Estimating Causal Quantile Effects}
  \author{Steven G. Xu, Shu Yang and Brian J. Reich\\
    Department of Statistics, North Carolina State University}
  \maketitle
} \fi

\if1\blind
{
  \bigskip
  \bigskip
  \bigskip
  \begin{center}
    {\LARGE\bf Title}
\end{center}
  \medskip
} \fi

\bigskip
\begin{abstract}
Standard causal inference characterizes treatment effect through averages, but the counterfactual distributions could be different in not only the central tendency but also spread and shape. To provide a comprehensive evaluation of treatment effects, we focus on estimating quantile treatment effects (QTEs). Existing methods that invert a nonsmooth estimator of the cumulative distribution functions 
forbid inference on probability density functions (PDFs), but PDFs can reveal more nuanced characteristics of the counterfactual distributions. We adopt a semiparametric conditional distribution regression model that allows inference on any functionals of counterfactual distributions, including PDFs and multiple QTEs. To account for the observational nature of the data and ensure an efficient model, we adjust for a double balancing score that augments the propensity score with individual covariates. We provide a Bayesian estimation framework that appropriately propagates modeling uncertainty. We show via simulations that the use of double balancing score for confounding adjustment improves performance over adjusting for any single score alone, and the proposed semiparametric model estimates QTEs more accurately than other semi/nonparametric methods. We apply the proposed method to the North Carolina birth weight dataset to analyze the effect of maternal smoking on infant's birth weight.
\end{abstract}

\noindent%
{\it Keywords:}  Quantile treatment effects; Double balancing score; Regression adjustment; Counterfactual distributions.
\vfill

\newpage
\spacingset{1.5} 
\section{Introduction}
\label{sec:intro}
Statistical methods for estimating causal effects of a treatment or exposure using observational data play a central role in many research fields, e.g., social and economic science \citep{imbens2015}, health care \citep{xie2020}, public policy \citep{frolich2013} and environmental science \citep{sun2021}. Most of the existing methods are cast in terms of the potential outcomes framework \citep{rubin1978}. Under this framework, causal inference can be drawn by comparing suitable functions of the distribution of \textit{counterfactual} or \textit{potential} outcomes, defined as the outcomes that would have been observed for the subjects if they were assigned treatment or control. 



Much of the causal inference literature to date focuses on examining the impact of a treatment on the central tendency of the counterfactual distributions, and most papers quantify such impact using the average treatment effect (ATE) or the conditional average treatment effect (CATE). In practice, however, asymmetric distributions are frequently encountered and are better summarized in quantiles. Furthermore, the counterfactual distributions under treatment and control could be different in not only central tendency but also spread or shape. In such cases, the ATE itself might not be enough or even fails to characterize the distributional differences. In contrast, comparing different parts of or the entire counterfactual distributions can provide more nuanced and valuable measures for exploring causal effects beyond a mean shift.

Quantile treatment effects (QTE) \citep{doksum1974} can capture heterogeneous causal effects of the treatment at different locations of the counterfactual distribution. They are defined as the difference between the counterfactual quantiles, marginalized over the distribution of confounders. Counterfactual quantiles can be estimated by either minimizing a weighted check loss or inverting an estimate of the cumulative distribution function (CDF) of the potential outcomes. In the latter case, the CDF can be naturally estimated using empirical expectation of thresholded outcomes, and therefore estimation of counterfactual quantiles largely reduces to estimation of counterfactual mean. A large number of existing works in QTE estimation consider estimators of these two types \citep[e.g.,][]{firpo2007,rothe2010,zhang2012,donald2014,yang2020,sun2021}. A drawback of these methods is that different quantiles have to be estimated separately, which could result in a non-monotonic estimate of the quantile function when the sample size is small. In addition, the discrete nature of these estimators forbids inference on the counterfactual probability density function (PDF). The PDF can reveal potentially interesting characteristics of the counterfactual distribution that cannot be revealed by the CDF, such as multimodality, and are often more visually interpretable to practitioners.

Density estimation is a statistically more complex problem than CDF estimation even in the non-counterfactual case. Consequently, counterfactual density estimation has received far less attention than its counterpart. Some early attempts are \citet{dinardo1996} and \citet{robins2001} which discussed possible estimators but did not proceed in full analysis. Recently, \citet{kim2018} adopted the kernel estimator from \citet{robins2001} and proposed a doubly robust like estimator that uses inverse probability weighting (IPW) for bias correction, but the resulting estimate is highly non-smooth. \citet{kennedy2021} proposed a similar estimation procedure, but uses a plug-in estimator based on truncated cosine series. However, extreme estimates of the propensity score (PS) as well as oscillating nature of the cosine series may result in an estimator with large variance. 

There has been a growing interest in adapting Bayesian semi/nonparametric models to flexibly model the counterfactual distribution and estimate associated causal effects. \citet{xu2018} proposed a two-stage approach for estimating the counterfactual distribution. First, the PS is modeled using probit regression and Bayesian additive regression trees \citep[BART;][]{chipman2010bart}. For each treatment
group, the counterfactual distribution conditional on the estimated PS is then modeled using a Dirichlet process mixture (DPM) of normals. Conditioning on the PS is an attractive solution to circumvent the curse of dimensionality while adjusting for confounding. However, the estimator might not be efficient enough when the outcome-PS model does not sufficiently characterize the outcome-covariate relationship.

In this paper, we propose a novel Bayesian semiparametric model for estimating the counterfactual distributions that allows inference on any functionals of the counterfactual distributions, including but not limited to counterfactual densities and quantile causal effects. Specifically, we model the counterfactual distributions by first modeling the conditional distributions of the outcomes given treatment and balancing score and then marginalizing it over the population distribution of the balancing score. Adjustment for balancing score is crucial in observational studies to reduce bias due to confounders. To formulate an efficient estimator that provides reliable inference of the counterfactual distributions, we propose to adjust for a double balancing score that augments the PS with individual covariates. To avoid the complexity due to posterior inference on the joint likelihood of the PS and the outcome \citep{zigler2013}, we adopt a sequential approach that separately estimates PS and the outcome in two stages. First, the PS is estimated using BART probit. Then, the balancing score containing posterior sample of the PS is used to estimate the outcome distribution. 
To ensure a flexible counterfactual density estimator that adapts to skewness, heavy-tailedness and multimodality, we extend the semiparametric quantile regression (SPQR) model by \citet{xu2021} to model the conditional outcome distribution using a finite mixture of shape-constrained splines, where the mixing weights are modeled by neural networks (NN). Observational studies often involve high-dimensional covariates. To improve the scalability of the proposed method and regularize its complexity, we give the NN weights Gaussian scale mixture (GSM) priors to automatically determine relevant features and encourage network sparsity. 


The proposed method differs from existing works in the following aspects. 
\begin{enumerate}
    \item Most existing works adjust for either the full-vector of covariates or the scalar PS \citep{imai2008,vansteelandt2014,xu2018}. Adjusting for covariates generally leads to smaller variability of outcome residual compared to adjusting for the PS. Adjusting for the PS alleviates the curse of dimensionality due to high-dimensional regression. The proposed double balancing score approach takes advantage of both approaches. It makes full use of the observed information by aggregating signals from both the covariates as well as the treatment assignment mechanism. The intuition is similar to that behind doubly robust methods. If the PS as a function of covariates does not summarize the outcome mean model well, the inclusion of individual covariates can help control the variability of outcome residual after adjusting for the PS. On the other hand, if the outcome-covariate relationship is complex and not well approximated, inclusion of the PS can reduce residual confounding after adjusting for individual covariates.
    \item Double score adjustment is different from doubly robust weighting methods that augments an IPW estimator with information from the outcome model \citep{zhang2012,kennedy2021} or doubly robust matching based on matching a double score that includes a PS and a prognostic score \citep{yang2020}. By incorporating the PS as a regressor, we alleviate the associated drawbacks of IPW such as high variance due to subjects with extreme PS and associated inefficiency of matching. 
    \item The estimation problem is substantially different to that in \citet{xu2021} where the focus is solely on conditional distribution and its quantiles. In the current context, the conditional distribution is an intermediate estimand used for estimating the marginal distribution and its quantiles. In addition, as we mention in Section~\ref{s:prior}, we account for the uncertainty of the distribution of covariates through Bayesian bootstrap \citep{rubin1981}.
    \item The proposed approach provides estimates of the full conditional counterfactual distributions given covariates which are unavailable from PS-based estimators \citep{xu2018}. These estimates have benefits for personalized medicine and allows one to assess the heterogeneity of treatment effect on individuals \citep{lu2018}.
\end{enumerate}

The rest of the paper is organized as follows. Section 2 defines the
causal estimands and states key identifying assumptions. Section 3 describes the double balancing score approach and the Bayesian semiparametric model to estimate counterfactual densities and QTEs. In Section 4, we compare the performance of the proposed approach with other counterfactual distribution estimators for a wide range of simulated data. We analyze the QTEs of smoking on low birth weight using the North Carolina birth weight data in Section 5 and conclude in Section 6.

\section{Preliminaries}
\label{sec:pre}
Let $T\in\cal T$ denote the treatment variable. For simplicity, we assume the treatment is binary, i.e., ${\cal T}=\{0,1\}$, such that $T=0$ is the control treatment and $T=1$ is the active treatment; the methods proposed in later sections extend naturally to multivalued treatments. Under the potential outcome framework, each level of treatment $t$ corresponds to a potential outcome $Y(t)$, representing the outcome that would have been observed for an individual if they were assigned treatment $T=t$. In real applications, $\{Y(0),Y(1)\}$ cannot be simultaneously observed, and only one corresponds to the actual outcome $Y$. Let $\bm{X}=(X_{1},...,X_{d})^{\top}\in\mathbb{R}^d$ denote a $d$-vector of exogenous covariates. The observed data consist of $(\bm{X}_i,T_i,Y_i)$, $i=1,...,n$, which we assume to be random samples of the joint distribution of $(\bm{X},T,Y)$. In addition, let $\pi(\bm{X})$ be the propensity score (PS),  defined as the probability of receiving active treatment given confounders, i.e., $\pi(\bm{X})=P(T = 1|\bm{X})$. 

To estimate the quantile causal effect, we make the following identifying assumptions.
\begin{assumption}[No-interference]The potential outcomes for one subject is independent of the treatment assignment of others.
\end{assumption}
\begin{assumption}[Consistency]For any subject, the observed outcome given the assigned treatment is equal to the potential outcome: $Y_i=T_iY_i(1)+(1-T_i)Y_i(0)$.
\end{assumption}
\begin{assumption}[Strongly ignorable treatment assignment, SITA]The treatment assignment is independent of the potential outcomes given the confounders: $\{Y(0),Y(1)\}\perp T|\bX$.
\end{assumption}
\begin{assumption}[Overlap]For any given values
of confounders, the probability of assignment to active
treatment is strictly between 0 and 1: $0< P(T=1|\bm{X}) < 1$.
\end{assumption}
The last two assumptions imply that the treatment assignment is independent of the potential outcomes given the PS, i.e., $\{Y(0),Y(1)\}\perp T|\pi(\bm{X})$.

Let $F_t(y)$ be the unconditional cumulative distribution function (CDF) of the potential outcome $Y(t)$, defined as
\begin{equation}\label{e:marginal}
    F_t(y) = \int F_t(y|\bm{x})dF(\bm{x}),\hspace{10pt} t=0,1,
\end{equation}
where $F_t(y|\bm{x})$ is the conditional CDF of $Y(t)$ given covariates $\bm{X}=\bm{x}$ and $F(\bm{x})$ is the CDF of $\bm{X}=\bm{x}$. Following Assumptions 3 and 4, $F_t(y)$ is identifiable through $F_t(y|\bm{X})$. We make the additional assumption that for $\tau\in(0,1)$, $F_t(y)$ is twice differentiable and strictly increasing so that its inverse function is well defined. Let $q_t(\tau)=F^{-1}_t(\tau)$ denote the $\tau$th marginal quantile of $Y(t)$. The $\tau$th QTE \citep{doksum1974,lehmann1975} is defined as
\begin{equation*}
    \Delta_{\text{QTE}}(\tau) = q_1(\tau) - q_0(\tau),
\end{equation*}
which can be intuitively interpreted as the ``horizontal distance" between the counterfactual distributions $F_1(y)$ and $F_0(y)$ in the target population. 


\section{Methodology}
\label{sec:meth}
\subsection{Double balancing score}
We propose to estimate the marginal quantiles $q_{t}(\tau)$ and the QTE by inverting a semiparametric estimator of the counterfactual distribution, $\hat{F}_t(y)$. Although estimating the counterfactual distribution is a considerably more challenging problem than directly modeling the quantile of interest, it avoids the necessity of fitting separate models for estimation of multiple quantiles. In addition, estimating the counterfactual distributions allows inference on not only quantiles but any functionals of the distributions. 

We adopt a balancing score approach for estimation of the counterfactual distribution. Let $\bm{S}$ denote a balancing score which is a function of covariates satisfying the condition $T\perp \bm{X}\mid\bm{S}$. Our method begins by approximating the conditional counterfactual distribution given the covariates, $F_t(y|\bm{X})$, with an estimator of the conditional distribution of the observed outcome given the treatment and balancing score, $\hat{F}(y|T,\bm{S})$. A flexible estimator of $F_t(y)$ can then be obtained by marginalizing $\hat{F}(y|T=t,\bm{S})$ over the distribution of the balancing score, $F(\bm{S})$, i.e.,
\begin{equation}\label{e:estimator}
    \hat{F}_t(y) = \int \hat{F}(y|t,\bm{s})dF(\bm{s}).
\end{equation}
Most of the existing works set the balancing score to be either the full covariate vector (outcome-covariate regression) or the PS (outcome-PS regression). We shall argue that both approaches suffer from some drawbacks. When the full covariate vector, $\bm{X}$, is used as the balancing score, Equation~\eqref{e:estimator} involves estimating a $(d+1)$-dimensional conditional distribution. In practice, $d$ is often large as $\bm{X}$ has to include as many potential confounders as possible in order to satisfy the \textbf{SITA} assumption. In addition, covariates that are strong predictors of the outcome are often included in $\bm{X}$ to help explain the variability of the outcome more comprehensively. Thus estimation of $F(y|T,\bm{X})$ requires high-dimensional nonparametric regression which is an inherently difficult problem, especially when the outcome and covariates exhibit a rather complex relationship. Using the PS, $\pi(\bm{X})$, as the balancing score alleviates the curse of dimensionality by collapsing the full covariate vector to a 
probability and simplifies the estimation problem to a low-dimensional one. However, since the true PS is unknown in practice, a first-stage model is required to approximate $\pi(\bm{X})$ using data. This will introduce additional model uncertainty and inflate the variance of the estimated counterfactual distribution. Furthermore, conditioning on $\pi(\bm{X})$ may result in loss of information when $\pi(\bm{X})$ is not sufficient to explain the outcome and covariate relationship. To alleviate their individual drawbacks and combine their strength, we propose the use of an augmented score $\bm{S}=\{\pi(\bm{X}),\bm{X}\}^{\top}$. Since both components of $\bm{S}$ are balancing scores, $\bm{S}$ is a ``double balancing score" \citep{hu2012}. The semiparametric double balancing score estimator of the counterfactual distribution is still \eqref{e:estimator} but uses $\bm{S}=\{\pi(\bm{X}),\bm{X}\}^{\top}$ instead of only $\pi(\bm{X})$ or $\bm{X}$. 

The proposed double balancing score estimation procedure has several advantages over existing methods for counterfactual distribution estimation. Compared to outcome-PS regression approaches \citep{xu2018}, the inclusion of the covariates makes $\bm{S}$ sufficient for explaining the outcome-covariate relationship, and therefore the resulting estimator may be more efficient. If the outcome-PS dependence relationship is complex but the outcome-covariate relationship is relatively simple, the proposed approach can benefit from the inclusion of covariates and estimate $F(y|T,\bm{X})$ more accurately. Compared to outcome-covariate regression approaches \citep{zhang2012,chernozhukov2013}, the inclusion of the PS incorporates signal from treatment assignment mechanism and utilizes the observed data fully. If the outcome-covariate relationship is complex but the outcome-PS dependence relationship is relatively simple, the proposed approach can benefit from the inclusion of PS and estimate $F(y|T,\bm{X})$ more accurately. The use of double balancing score is not the only way to aggregate information from both the covariates and the PS. Instead of including $\pi(\bm{X})$ as a regressor, one can use IPW to adjust for the bias of an initial conditional counterfactual distribution estimator \citep[e.g.,][]{kim2018,kennedy2021}. However, the advantage of incorporating $\pi(\bm{X})$ as one component of $\bm{S}$ is that associated drawbacks of IPW such as high
variance due to subjects with extreme PS can be reduced.

\subsection{Semiparametric counterfactual distribution estimation}

As shown in \eqref{e:estimator}, estimating $F_t(y)$ boils down to approximating the nonparametric conditional outcome CDF $F(y|T,\bm{S})$, or equivalently, the conditional outcome PDF $f(y|T,\bm{S})$. Without loss of generality, we assume that $Y$ lies in the unit interval. This can be achieved by transforming $Y$ from its original support using an appropriate monotone mapping. To allow a flexible model for the counterfactual density that can adapt to
skewness, heavy-tailedness and multimodality, we extend the SPQR model by \citet{xu2021} to model the conditional distribution of outcomes given the treatment and balancing score. Let $M_1(y),...,M_K(y)$ be $K$ second-order M-spline basis functions with equally-spaced knots spanning $[0,1]$, and let $I_1(y),...,I_K(y)$ be second-order I-spline basis functions with the same knots. We model the conditional PDF and CDF of $Y$ respectively by
\begin{equation}\label{e:isp}
    f(y|T,\bm{S}) = \sum_{k=1}^K\theta_k(T,\bm{S})M_k(y)\mbox{\ \ \ and\ \ \ } F(y|T,\bm{S}) = \sum_{k=1}^K\theta_k(T,\bm{S})I_k(y)
\end{equation}
where the mixture weights $\theta_k(T,\bm{S})$ satisfy $\theta_k(T,\bm{S})\ge 0$ and $\sum_{k=1}^K\theta_k(T,\bm{S})=1$ for all possible $T$ and $\bm{X}$. 

The weights are then modeled using feed-forward neural networks (NN) with $L$ layers ($L-1$ hidden layers) and a softmax output activation. Let $\mathcal{W}=\{\bW^{(1)}, ...,\bW^{(L)}\}$ denote the set of weight matrices with $\bW^{(l)}\in\mathbb{R}^{(V_{l}+1)\times V_{l-1}}$, where $V_l$ is the number of units (excluding the intercept/bias node) in layer $l$ and $V_0=d+2$. Denote $\hat{\theta}_k(T,\bm{S},\mathcal{W})$ as the NN estimator for $\theta_k(T,\bm{S})$ with modeling parameters $\mathcal{W}$, then $\hat{\theta}_k(T,\bm{S},\mathcal{W})$ can be expressed hierarchically as
\begin{equation}\label{e:nn}
\begin{split}
    \hat{\theta}_k(T,\bm{S},\mathcal{W}) &= \mbox{softmax}\lcbk z_k^{(L)}(T,\bm{S},\mathcal{W})\rcbk,\\
    z_k^{(l)}(T,\bm{S},\mathcal{W}) &= W^{(l)}_{k0}+\sum_{j=1}^{V_{l-1}}W^{(l)}_{kj}\phi\lcbk z_j^{(l-1)}(T,\bm{S},\mathcal{W})\rcbk\mbox{\ \ \ for\ \ } l = 2,...,L,\\
    z_k^{(1)}(T,\bm{S},\mathcal{W})&=W^{(1)}_{k0}+W^{(1)}_{k1}T+\sum_{j=1}^{d+1}W^{(1)}_{k(j+1)}S_j,
\end{split}
\end{equation}
where $\mbox{softmax}(u_k)=e^{u_k}/\sum_{k=1}^Ke^{u_k}$, $\phi(\cdot)$ is the hidden layer activation, and $S_j$ is the $j$th component of $\bm{S}$. Throughout the paper, we assume $\phi(\cdot)$ to be the hyperbolic tangent function, i.e., $\phi(u)=(e^{2u}-1)/(e^{2u}+1)$. 

The NN is known to scale well with the dimension of the regressors, and thus the proposed estimator is particularly suitable for applications in which the observed covariates contain both confounders as well as predictors for only the outcome. The conditional distribution estimator combining \eqref{e:isp} and \eqref{e:nn} is extremely flexible. In fact, it has been shown that SPQR can approximate any smooth conditional distributions when $\bm{S}=\bm{X}$ \citep{xu2021}. The smoothness of regression splines also ensures that the quantiles and the causal effects are well defined.

\subsection{Bayesian estimation of quantile causal effects}\label{s:prior}
To allow uncertainty quantification of the estimated counterfactual distributions and quantile causal effects. We adopt a Bayesian framework for estimating the counterfactual distributions. A fully Bayesian approach using a PS requires posterior inference of the joint likelihood of the PS and outcome \citep{zigler2013}, which may be complex especially when the PS is modeled nonparametrically. To simplify the problem, we adopt a sequential approach and estimate $\pi(\bm{X})$ and $F(y|T,\bm{S})$ separately in two stages. We use a fully Bayesian approach to model and make inference on $\pi(\bm{X})$ using BART probit, $F(y|T,\bm{S})$ using Bayesian NN, and $F(\bm{S})$ using Bayesian bootstrap \citep{rubin1981}.

\subsubsection*{Propensity score}
Since $\pi(\bm{X})$ is the conditional expectation of a binary variable, it is natural to use logistic regression to investigate the dependence relationship. However, parametric models are prone to misspecification when the relationship between the treatment and confounders is highly nonlinear. To ensure flexible estimation of the PS, we follow \citet{xu2018} and model $\pi(\bm{X})$ using BART probit \citep{chipman2010bart}, $\hat{\pi}(\bm{X}) = \Phi\lcbk\sum_{j=1}^m\tilde{T}_j(\bm{X})\rcbk$, where $\Phi(\cdot)$ is the standard normal CDF and $\tilde{T}(\cdot)$ is a regression tree model. Let $\lcbk \hat{\pi}^{(k)}(\bm{X}_i),\ k=1,...,N_{\pi}\rcbk$ denote $N_{\pi}$ posterior samples of $\hat{\pi}(\bm{X}_i)$, $i=1,...,n$. We propagate uncertainty of the PS into the model of $F(y|T_i,\bm{S}_i)$ by treating $\lcbk \bm{S}^{(k)}_i,\ k=1,...,N_{\pi}\rcbk$ as informative priors for $\bm{S}_i$, where $\bm{S}^{(k)}_i=\lcbk\hat{\pi}^{(k)}(\bm{X}_i),\bm{X}_i\rcbk^{\top}$. We then estimate $F(y|T_i,\bm{S}^{(k)}_i)$ using \eqref{e:nn} for each $k$. 

\subsubsection*{Priors over NN weights}
To complete the Bayesian formulation for the conditional distribution estimator, we must assign priors to the NN weights in SPQR. The uncertainty of $f(y|T,\bm{S})$ and $F(y|T,\bm{S})$ can then be characterized by the posterior distributions of $\hat{\theta}_k(T,\bm{S},\mathcal{W})$. Zero-mean Gaussian distributions are widely used as priors for Bayesian NN weights owning to their \textit{weight-decay} property.
They have been explored both in classic works on shallow NNs \citep{neal1996,mackay1992} as well as more recent works on deep NNs \citep{matthews2018,fortuin2021}. The performance of Gaussian priors depends heavily on the variance hyper-prior structure which allows one to encode problem-specific
beliefs as well as general properties about weights. An isotropic Gaussian prior that assigns all weights a common variance, although a convenient choice, can in fact lead to inflated predictive uncertainties \citep{neal1996}. 

Gaussian scale mixture (GSM) priors have been shown to be effective for Bayesian inference of large NNs \citep[e.g.,][]{cui2021informative}. For each layer $l\in\{1,...,L\}$, the GSM prior on $W_{kj}^{(l)}$ can be expressed hierarchically as
\begin{equation}\label{e:priors}
    W_{kj}^{(l)}|\kappa^{(l)},\omega_{j}^{(l)} \sim \mathcal{N}\left(0,\frac{1}{\kappa^{(l)}\omega^{(l)}_j}\right), \ p(\omega_j^{(l)})\sim p(\omega_j^{(l)};\gamma_{\lambda})\mbox{\ \ \ for\ \ } j\ge 0,
\end{equation}
where $\kappa^{(l)}$ is the layer-wise global precision shared by all weights in layer $l$, which can either
be set to a constant value or estimated using non-informative priors, and $\omega_{j}^{(l)}$ is a unit-wise local precision with hyper-prior $p(\omega_j^{(l)};\gamma_\omega)$. Although many more advanced settings are equally applicable \citep[e.g.,][]{ghosh2019}, we use Gamma distributions as hyper-priors for both the global precision $\kappa^{(l)}$ and the local precision $\omega_{j}^{(l)}$, i.e., 
\begin{equation*}\label{e:igamma}
    p(\kappa^{(l)};\gamma_{\kappa})=\mathcal{G}amma(a_{\kappa},b_{\kappa})  \mbox{\ \ \ and\ \ \ }
    p(\omega_j^{(l)};\gamma_{\omega})=\mathcal{G}amma(a_{\omega},b_{\omega}),
\end{equation*}
as it simplifies the sampling algorithm greatly. 
By assigning all outgoing weights $W_{kj}^{(l)}$ from node $j$ in layer $l$ a common scale parameter, the GSM prior achieves \textit{Automatic Relevance Determination} (ARD) \citep{mackay1992} which allows weights that are associated with relevant features to be large and forces weights that are associated with irrelevant features to be small. The ARD property is especially appealing to the current context since $f(y|T,\bm{S})$ is often high-dimensional to satisfy \textbf{SITA}, and high-dimensional conditional distributions often have a \textit{sparse} structure \citep{izbicki2016}. In addition, the hierarchical structure of GSM prior allows characterization of heavy-tailed weight distributions which are often observed in deep NNs \citep{fortuin2021}. 

To sample the parameters $W^{(l)}_{ij}$, $\kappa^{(l)}$ and $\omega^{(l)}_j$ in \eqref{e:priors}, we use Markov chain Monte Carlo (MCMC) methods to approximate their posterior distributions. Specifically, we follow the strategy of \citet{neal1996} and use a block-updating scheme that combines the strength of two MCMC algorithms. The posterior distribution of the weight parameters $W_{kj}^{(l)}$ is high-dimensional and has a complex geometry
\begin{equation*}
p(\mathcal{W};\kappa^{(l)},\omega^{(l)}_j)\propto\prod_{l=1}^L\prod_{k=1}^K\prod_{j=0}^{V_{l-1}}\mathcal{N}\lpth W_{kj}^{(l)}|0,\frac{1}{\kappa^{(l)}\omega^{(l)}_j}\rpth\times\prod_{i=1}^n f(y_i|T_i,\bm{S}_i,\mathcal{W}),
\end{equation*}
therefore it is approximated using the no-U-turn sampler \citep[NUTS,][]{hoffman2014}. The prior distributions of the precision hyperparameters $\kappa^{(l)}$ and $\omega_{j}^{(l)}$ are conjugate, therefore their full conditional distributions can be derived analytically as
\begin{equation*}
    \begin{split}
        \kappa^{(l)}|W^{(l)}_{kj},\omega^{(l)}_j&\sim\mathcal{G}amma\lpth a_\sigma+\frac{V^{(l+1)}(V^{(l)}+1)}{2},b_\sigma+\frac{\sum_{k=1}^K\sum_{j=0}^{V_{l-1}}\lpth \sqrt{\omega^{(l)}_j}W^{(l)}_{kj}\rpth^2}{2}\rpth\\
        \omega_j^{(l)}|W^{(l)}_{kj},\kappa^{(l)}&\sim\mathcal{G}amma\lpth a_\lambda+\frac{V^{(l+1)}}{2},b_\lambda+\frac{\sum_{k=1}^K\lpth \sqrt{\kappa^{(l)}}W^{(l)}_{kj}\rpth^2}{2}\rpth
    \end{split}
\end{equation*}
and their posterior distributions can be approximated using Gibbs sampling \citep{geman1984}. 

\subsubsection*{Bayesian bootstrap}

Finally, to estimate the unconditional counterfactual distribution, we need to marginalize the estimated conditional distribution of outcomes given the treatment and the balancing score over the population distribution of the balancing score. The population distribution of the covariates (and the balancing score) is typically not known in observational studies. Under the assumption that the observed group of individuals is a simple random sample from the target population, a reasonable estimate of the covariate distribution is the empirical distribution $H_n(\bm{X})=\frac{1}{n}\sum_{i=1}^n\delta_{\bm{X}_i}$ where $\delta_{\bm{X}_i}$ is a degenerate distribution at $\bm{X}_i$. Similarly, for each posterior sample of the balancing score, $\bm{S}^{(k)}$, we can estimate its distribution by $H_n^{(k)}(\bm{S})=\frac{1}{n}\sum_{i=1}^n\delta_{\bm{S}^{(k)}_i}$. To incorporate uncertainty of this distribution, we follow the suggestion of \citet{xu2018} and use the Bayesian bootstrap \citep{rubin1981}. Consider the weighted empirical distribution model $H_n^{(k)}(\bm{S},\bm{u})=\sum_{i=1}^nu_i\delta_{\bm{S}^{(k)}_i}$ where the sampling weights $u_i$ are unknown, non-negative, and sum to one. We give the weights non-informative priors $p(\bm{u})\propto\prod_{i=1}^nu_i^{-1}$. The uncertainty of the balancing score distribution can then be characterized by the posterior distribution of $\bm{u}$ which follows $\mathcal{D}irichlet(1,...,1)$. 

The steps for estimating the counterfactual distribution and QTE are summarized in Algorithm~\ref{a:qte}. After obtaining the posterior samples: $f^{(jl)}_0(y)$, $f^{(jl)}_1(y)$ and $\Delta^{(jl)}_{\text{QTE}}(\tau)$, for $j=1,...,N_\pi$ and $l=1,...,N_{\mathcal{W}}$. We can compute the estimated counterfactual densities and QTE by averaging the samples
\begin{equation*}
\begin{split}
    \hat{f}_t(y) &= \frac{1}{N_\pi}\frac{1}{N_{\mathcal{W}}}\sum_{j=1}^{N_\pi}\sum_{l=1}^{N_{\mathcal{W}}}f^{(jl)}_t(y)\mbox{\ \ \ for\ \ }t=0,1\\ 
    \widehat{\Delta}_{\text{QTE}}(\tau) &= \frac{1}{N_\pi}\frac{1}{N_{\mathcal{W}}}\sum_{j=1}^{N_\pi}\sum_{l=1}^{N_{\mathcal{W}}} \Delta^{(jl)}_{\text{QTE}}(\tau)
\end{split}
\end{equation*}
and compute credible intervals (CIs) using relevant percentiles. 
\begin{algorithm}
\caption{Bayesian semiparametric estimation of counterfactual densities and QTE}\label{a:qte}
\textbf{Input:} Observed data $(\bm{X}_i,T_i,Y_i)$\\
\textbf{Output:} Posterior samples of $f_0(y)$, $f_1(y)$ and $\Delta_{\text{QTE}}(\tau)$
\begin{algorithmic}[1]
\State Sample $\hat{\pi}^{(j)}(\bm{X}_i)$, $j=1,...,N_\pi$, from BART probit 
\For{$j=1,...,N_\pi$}
\State $\bm{S}_i^{(j)}\gets \lcbk\hat{\pi}^{(j)}(\bm{X}_i),\bm{X}_i\rcbk^{\top}$
\State Sample $\mathcal{W}^{(jl)}$, $l=1,...,N_{\mathcal{W}}$, from Bayesian NN\Comment{Equation~\eqref{e:priors}}
\For{$l=1,...,N_{\mathcal{W}}$}
\State Sample $u^{(jl)}_i$ from $\mathcal{D}irichlet(1,...,1)$; $H^{(jl)}_n(\bm{S},\bm{u})=\sum_{i=1}^nu^{(jl)}_i\delta_{\bm{S}^{(j)}_i}$
\For{$t=0,1$}
\State Compute\Comment{Equation~\eqref{e:isp}}
\begin{equation*}
    \begin{split}
        F^{(jl)}(y|T=t,\bm{S}_i^{(j)})&=\sum_{k=1}^K\hat{\theta}_k\lpth t,\bm{S}_i^{(j)},\mathcal{W}^{(jl)}\rpth I_k(y)\\
        f^{(jl)}(y|T=t,\bm{S}_i^{(j)})&=\sum_{k=1}^K\hat{\theta}_k\lpth t,\bm{S}_i^{(j)},\mathcal{W}^{(jl)}\rpth M_k(y)
    \end{split}
\end{equation*}
\State Compute \Comment{Equation~\eqref{e:estimator}}
\begin{equation*}
    \begin{split}
        F^{(jl)}_t(y)&=\int F^{(jl)}(\tilde{y}_{g}|T=t,\bm{s}^{(j)})dH^{(jl)}_n(\bm{s}^{(j)})\\
        &=\sum_{i=1}^nu^{(jl)}_i F^{(jl)}(\tilde{y}_{g}|T=t,\bm{S}_i^{(j)})\\
        f^{(jl)}_t(y)&=\sum_{i=1}^nu^{(jl)}_i f^{(jl)}(\tilde{y}_{g}|T=t,\bm{S}_i^{(j)})
    \end{split}
\end{equation*}
\State $q_t^{(jl)}(\tau)\gets y$ s.t. $F_t(y)=\tau$
\EndFor
\State Compute $\Delta^{(jl)}_{\text{QTE}}(\tau)=q_1^{(jl)}(\tau)-q_0^{(jl)}(\tau)$ 
\EndFor
\EndFor
\State \textbf{return} $f^{(jl)}_0(y)$, $f^{(jl)}_1(y)$ and $\Delta^{(jl)}_{\text{QTE}}(\tau)$, $j=1,...,N_\pi$ and $l=1,...,N_{\mathcal{W}}$ 
\end{algorithmic}
\end{algorithm}

\section{Simulation}

We examine the performance of the proposed approach in estimating counterfactual densities and QTE using four simulations. The details for each simulation are provided as follows. (1) Simulation 1 explores the effect of adjusting for individual covariates in addition to the PS in presence of nonconfounding covariates. (2) Simulation 2 explores the effect of adjusting for the PS in addition to the covariates when the outcome-covariate relationship is complex but the outcome-PS relationship is simple. (3) Simulation 3 explores the effect of adjusting for the covariates in addition to the PS when the outcome-PS relationship is complex but the outcome-covariate relationship is simple. (4) Simulation 4 investigates the performance of the proposed approach when the conterfactual distributions are strongly non-Gaussian. For 
all four simulations, the proposed approach, which we call SPQR-DS (double score), is compared with two counterfactual distribution estimators: the DPM-BART estimator proposed by \citet{xu2018} and the truncated series (TS) estimator proposed by \citet{kennedy2021}. The DPM-BART estimator uses outcome-PS regression. First, the PS is modeled flexibly using BART probit \citep{chipman2010bart}. For each treatment group, the distribution of potential outcomes conditional on the estimated PS is estimated using a Dirichlet process mixture (DPM) of normals and finally marginalized over the population distribution of the covariates. Similar to the proposed approach, the DPM estimator incorporate uncertainty of the PS through its posterior samples and the uncertainty of the covariate distribution through Bayesian bootstrap. The TS estimator uses a doubly robust like approach. First, the distribution of the potential outcomes conditional on the full covariate vector is estimated using the kernel-smoothed approach of \citet{kim2018}. An intial estimate of the counterfactual density is then obtained by projecting the nonparametric kernel estimator to a truncated cosine series. Finally, IPW is combined with this initial estimate for bias correction. In addition to the three main approaches described above, we also compared DPM for modeling the entire distribution of the outcome
given covariates (DPM-X), and the proposed estimator that uses either only the individual covariates (SPQR-X) or only the PS (SQRT-BART). We did not compare with DPM that uses double balancing score because the perfect multicollinearity renders DPM numerically unstable.

To estimate the PS, we sample the posterior distribution of the parameters in the BART probit model using the R package \textit{BayesTree} with default priors \citep{chipman2010bart}. We run the MCMC for 1000 iterations and save every 100th iteration after discarding 500 iterations as burn-in. To estimate the conditional distribution using SPQR, we use the R package \textit{SPQR} with the priors described in Section~\ref{s:prior}. We set $a_\kappa=b_\kappa=a_\omega=b_\omega=0.01$ to give the variance parameters uninformative priors. We model the mixture weights using NNs with one hidden layer and select the number of basis functions and hidden neurons from $K=\{8,10,12\}$, $V_1=\{5,8,10\}$ using WAIC \citep{watanabe2013}. We run the MCMC for 3000 iterations and save every 10th iteration after discarding 1000 iterations as burn-in. To estimate the conditional distribution using DPM of normals, we use the R code in \citet{xu2018}. Following the original authors' setup, we run the MCMC for 900 iterations save every 2nd iteration after discarding 500 iterations as burn-in. For TS, we use the R package \textit{npcasual} to fit the model and its built-in cross-validation (CV) routine for basis selection. For each level of treatment, we consider up to 10 basis terms and choose the best model using 5-fold CV. During our experiment, we found that the algorithm used to fit the TS model is sometimes numerically unstable due to multiple uses of numerical integration. In cases when CV fails, we use the default setting of five basis functions for both treatment groups.

Each simulation design generates covariates $\bm{X}$, then binary treatments $T|\bm{X}$, and finally potential outcomes $Y(0)$ and $Y(1)$ for each counterfactual regime. The observed response is then $Y = TY(1)+(1-T)Y(0)$. We generate 100 replicated datasets with sample size $n=500$ for each design. We compare the estimated density of potential outcomes from the different approaches. In particular, we use the integrated squared error (ISE) to quantify the difference between an estimated density and its ground truth. We estimate the ISE by
\begin{equation*}
\begin{split}
    \mbox{ISE}(\hat{f},f)&\equiv\int\lsbk\hat{f}(y)-f(y)\rsbk^2dy\\
    &\approx(g_1-g_0)\sum_{i=1}^{n_\text{grid}-1}\lsbk\hat{f}(g_i)-f(g_i)\rsbk^2+\frac{\lsbk\hat{f}(g_0)-f(g_0)\rsbk^2+\lsbk\hat{f}(g_{n_\text{grid}})-f(g_{n_\text{grid}})\rsbk^2}{2}
\end{split}
\end{equation*}
where $\hat{f}(y)$ is the estimated density, $f(y)$ is the true density, $g_i$ are equidistant grid points and $n_\text{grid}=200$ is used in the simulations. We calculate QTE for 19 quantiles ($\tau=0.05,0.10,...,0.95$). We compare all the different approaches in terms of $\tau$-specific root mean squared error (RMSE) and average absolute bias (AAB)
\begin{equation*}
    \begin{split}
        \mbox{RMSE}(\tau) &= \frac{1}{100}\sum_{i=1}^{100}\lsbk\widehat{\Delta}^{(i)}_\text{QTE}(\tau)-\Delta_\text{QTE}(\tau)\rsbk\\
        \mbox{AAB}&=\frac{1}{19}\sum_{i=1}^{19}\left|\widehat{\Delta}_\text{QTE}(\tau_i)-\Delta_\text{QTE}(\tau_i)\right|
    \end{split}
\end{equation*}
where $\widehat{\Delta}^{(i)}_\text{QTE}(\tau)$ denote the estimated $\tau$-QTE from the $i$th replicated data set.

\subsection{Simulation 1}
Each subject is associated with 5 continuous covariates of which $J$ are confounders and $5-J$ are nonconfounding covariates. The true PS model is a logistic regression model that includes only main effects of the confounders. One
potential outcome is a mixture of normal models and the
other has a skewed distribution for the error term. The exact form of the true model is 
\begin{equation*}
    \begin{split}
        X_j&\sim \mathcal{U}(-2,2)\hspace{5mm} j=1,...,5\\
        T|\bm{X}&\sim \mathcal{B}ern\lpth\mbox{expit}\lpth\frac{4}{J}\sum_{j=1}^JX_j\rpth\rpth\\
        Y(0)|\bm{X}&=-2.3+Z_1+Z_1^2+\epsilon\\
        &\ \epsilon\sim 0.75\mathcal{N}^{+}(0,0.9^2)+0.25\mathcal{N}^{-}(0,0.3^2)\\
        Y(1)|\bm{X}&\sim 0.7\mathcal{N}\lpth-2.5+5Z_2,0.35^2\rpth+0.3\mathcal{N}\lpth2.5-5Z_2,0.35^2\rpth
    \end{split}
\end{equation*}
where $Z_k = \mbox{expit}(0.8\sum_{j=1}^5X_j+0.1\sum_{j=1}^5|X_j|^k)$. We experiment with two settings of $J$: $J=0$ such that there is no confounding and the data represent observations from a randomized trial, and $J=2$ such that there is strong confounding. In both cases, nonconfounding covariates contribute to variability of outcome residual after adjusting for PS. 

\begin{figure}[!tbh]
\centerline{\includegraphics[width=0.8\linewidth]{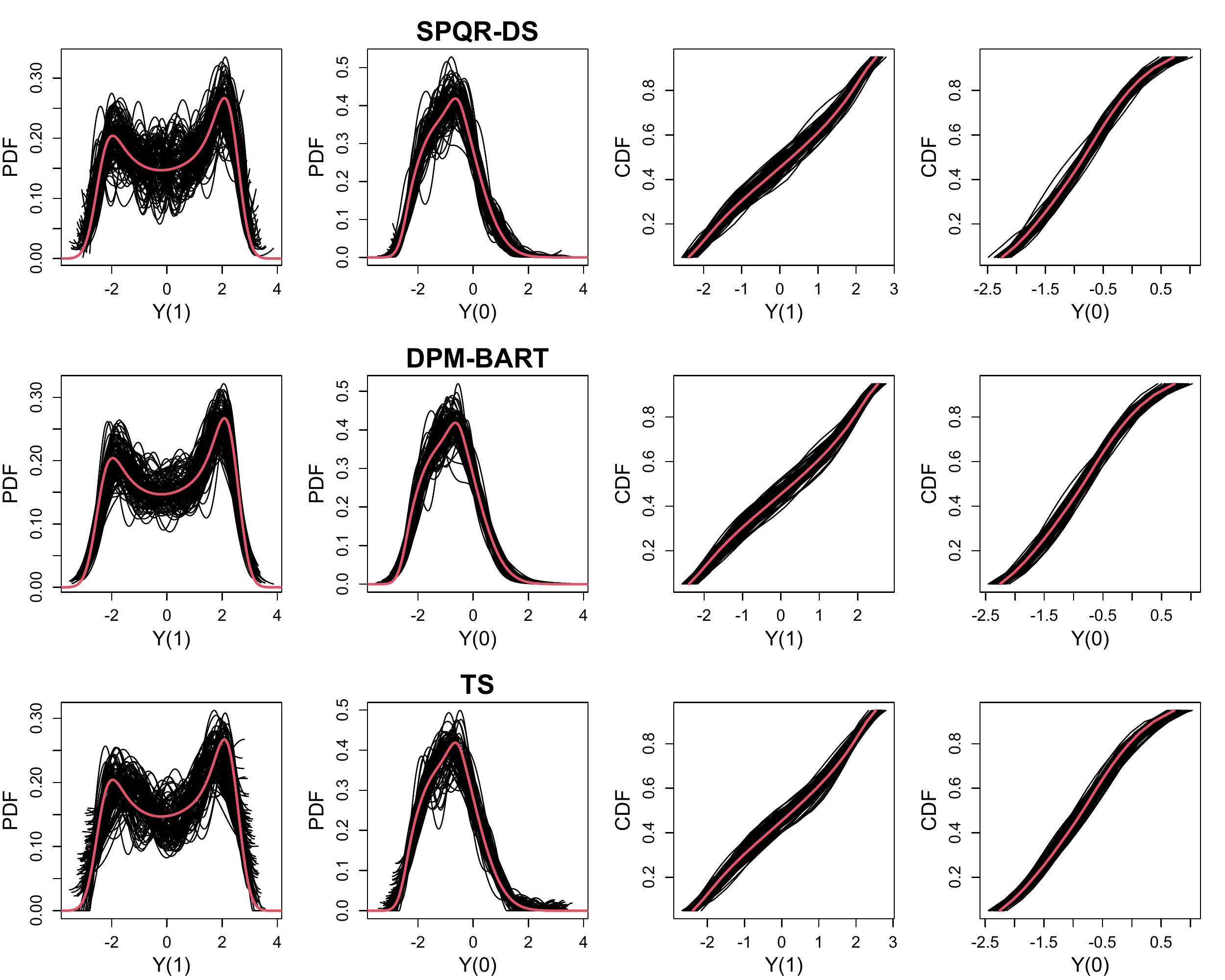}}
\caption{Simulation 1, $J=0$: Estimated PDFs and CDFs (black lines) of potential outcomes compared to the ground truth (red line), for 100 replicates for the SPQR-DS approach, the DPM-BART approach, and the TS approach.}\label{f:ex1_0}
\end{figure}

\begin{figure}[tbh]
\centerline{\includegraphics[width=\linewidth]{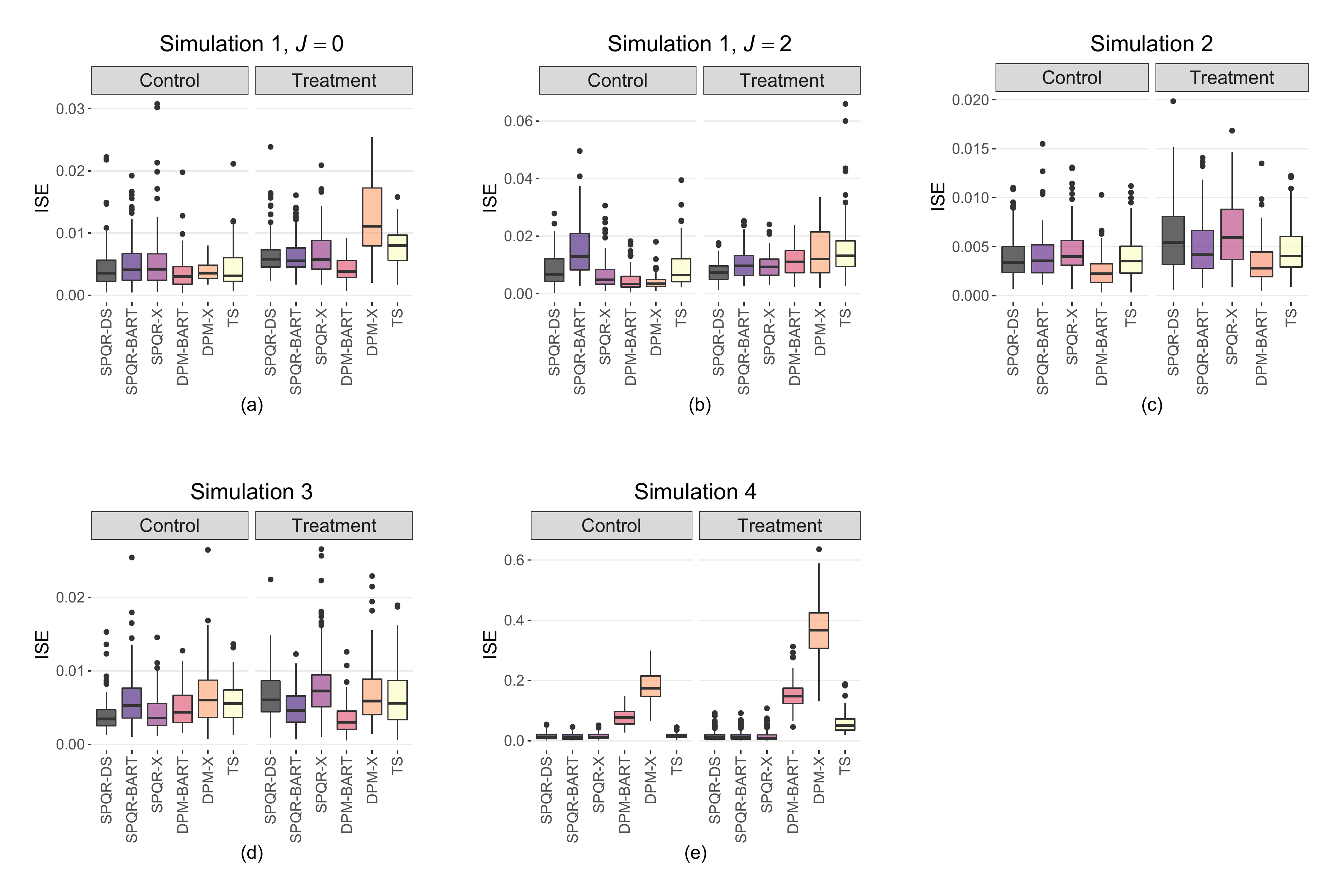}}
\caption{Simulation results. ISE of estimated counterfactual densities for 100 replicates for all approaches.}\label{f:ex_ise}
\end{figure}

\begin{figure}[tbh]
\centerline{\includegraphics[width=0.8\linewidth]{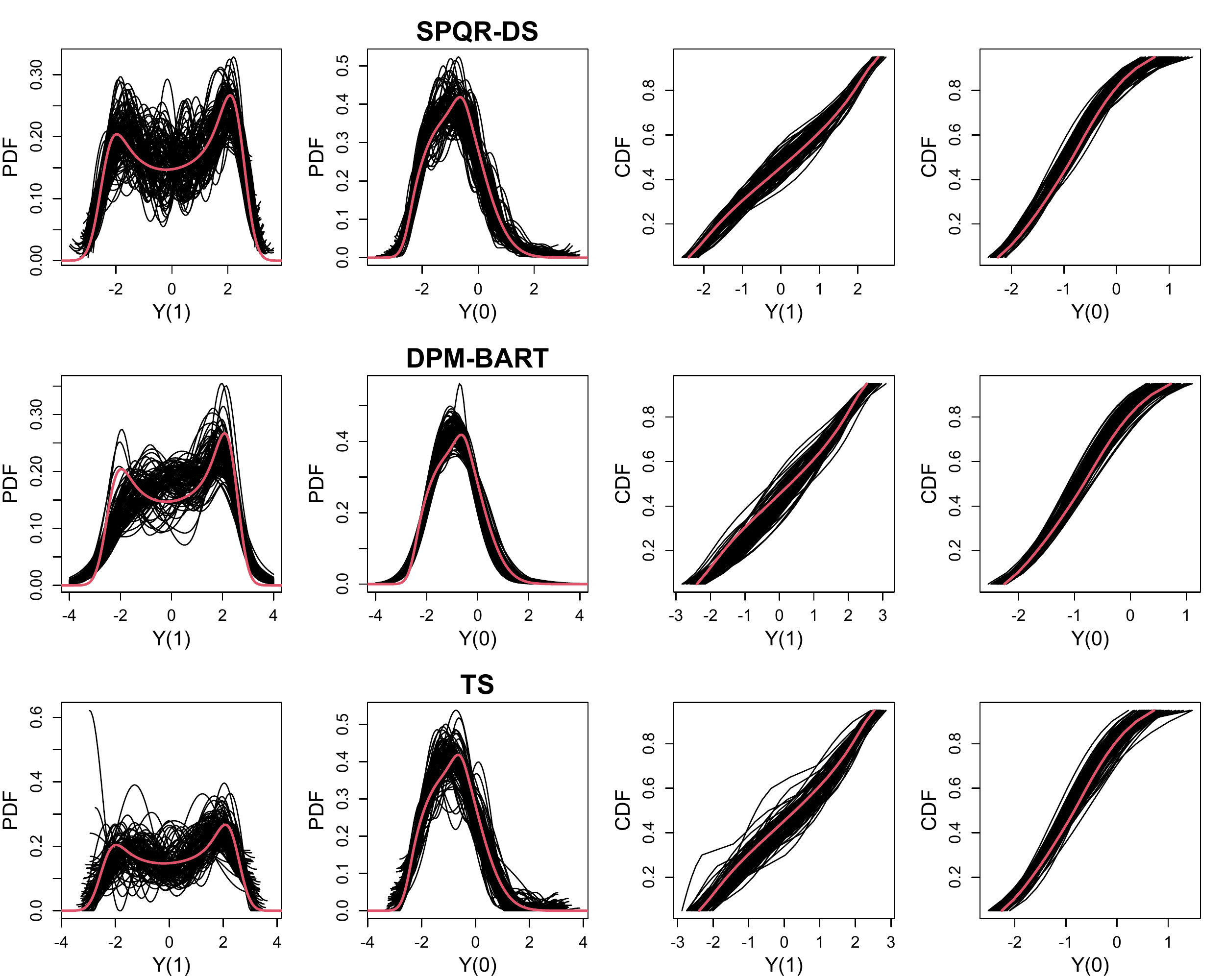}}
\caption{Simulation 1, $J=2$: Estimated PDFs and CDFs (black lines) of potential outcomes compared to the ground truth (red line), for 100 replicates for the SPQR-DS approach, the DPM-BART approach, and the TS approach.}\label{f:ex1_2}
\end{figure}

In the case of $J=0$, Figure~\ref{f:ex1_0} displays the true and estimated counterfactual PDFs and CDFs from the 100 replicates for SPQR-DS, DPM-BART and TS. The results show that all three approaches correctly estimate the bimodal and skewed distributions of the potential outcomes. However, compared to SPQR-DS and DPM-BART, the TS model has significantly larger variance in the tails of the distribution. Figure~\ref{f:ex_ise}(a) displays the boxplots of ISE for all approaches. The results show that despite only using the PS, the DPM-BART approach yields the best performance in density estimation. One possible explanation might be that when there is no confounding, $f_t(y)$ does not require identification through $f_t(y|\bm{X})$ Therefore the normal outcome distributions can be easily approximated by DPM since it is the correct model. On the other hand, directly estimating the full conditional using DPM yields poor performance for the control density, suggesting that DPM may not scale well with the dimension of the conditioning variables. The proposed approach seems to benefit from the inclusion of both the PS and the full covariate, yielding smaller variance than its two sub-models. The same plots in the case of $J=2$ are shown in Figure~\ref{f:ex1_2} and \ref{f:ex_ise}(b). The results show that even in presence of strong confounding, the proposed approach still yields competitive accuracy for counterfactual distributions. The DPM-BART approach yields biased estimates for the treatment distribution. The TS approach is generally accurate, but is unstable and yields more outlying estimates. Table~\ref{t:sim1} displays the RMSE of QTEs at five selected quantile levels and AAB based on all 19 quantile levels for all approaches. The results show that the proposed model yields the best QTE estimation in both cases. The use of double-balancing score for residual adjustment is particularly advantageous in presence of strong confounding, reducing the bias and improving the efficiency of the estimator.

\begin{table}[tb]
    \small
    \centering
    \caption{Simulation 1. AAB and $\tau$-specific RMSE of QTE for all approaches. AAB is calculated using all 19 quantiles, and standard deviation is given in parentheses.}
    \label{t:sim1}
    \midsepremove
    \begin{tabular}{@{}*{8}c@{}}\toprule
     & & \multicolumn{6}{c}{RMSE of $\widehat{\Delta}_{\text{QTE}}(\tau)$}\\ 
    \cmidrule(lr){3-8}
     & $\tau$ & SPQR-DS  & SPQR-BART & SPQR-X & DPM-BART  & DPM-X & TS\\
    \midrule
    \multicolumn{2}{l}{$J=0$}\\
                       &  0.1   & 0.10 & 0.09 & 0.11 & 0.10 & 0.12 & 0.11 \\
                       &  0.25   & 0.16 & 0.16 & 0.16 & 0.16 & 0.21 & 0.14 \\
                       &  0.5   & 0.21 & 0.22 & 0.21 & 0.20 & 0.15 & 0.21 \\
                       &  0.75   & 0.12 & 0.15 & 0.12 & 0.12 & 0.23 & 0.15 \\
                       &  0.9   & 0.11 & 0.13 & 0.13 & 0.11 & 0.12 & 0.11 \\
    \cmidrule(lr){2-8}                  
    & AAB &  \textbf{0.12 (0.06)} & 0.13 (0.07) & \textbf{0.12 (0.06)} & \textbf{0.12 (0.06)} & 0.16 (0.04) & \textbf{0.12 (0.05)} \\
    \multicolumn{2}{l}{$J=2$}\\
                       &  0.1   & 0.15 & 0.16 & 0.17 & 0.24 & 0.19 & 0.24 \\
                       &  0.25   & 0.21 & 0.24 & 0.21 & 0.44 & 0.28 & 0.34 \\
                       &  0.5   & 0.24 & 0.30 & 0.30 & 0.28 & 0.24 & 0.36 \\
                       &  0.75   & 0.17 & 0.20 & 0.23 & 0.16 & 0.29 & 0.21 \\
                       &  0.9   & 0.19 & 0.22 & 0.21 & 0.21 & 0.17 & 0.20 \\ 
    \cmidrule(lr){2-8}                  
    & AAB &  \textbf{0.17 (0.06)} & 0.19 (0.10) & 0.19 (0.08) & 0.23 (0.10) & 0.20 (0.08) & 0.23 (0.10)\\ \bottomrule
    \end{tabular}
    \begin{tablenotes}
        \footnotesize
        \item[a] Note: best-performing models with the smallest average AAB are in bold fonts.
    \end{tablenotes}
    \midsepdefault
\end{table}

\subsection{Simulation 2}
Each subject is associated with 6 continuous and 6 binary confounders. The true PS model is a logistic regression model that includes interactions between the confounders. Both potential outcomes follow mixture of normal models with the normal distributions depending on functions of the PS. In addition, the normal mixture model under control has weights depending on the PS. The exact form of the true model is given in Appendix A.

Figure~\ref{f:ex2} displays the true and
estimated counterfactual PDFs and CDFs from the
100 replicates for SPQR-DS, DPM-BART and TS. The results show that all three approaches correctly estimate the counterfactual distributions. Figure~\ref{f:ex_ise}(c) displays the boxplots of ISE for all approaches except DPM since the algorithm failed on all 100 replicates. This suggests that DPM is generally not suitable for high-dimensional conditional distribution regression. The results show that all approaches that incorporate information of PS perform better than those who do not (in this case only SPQR-X). This is expected since the outcome-PS dependence relationship has a simple form. The DPM-BART approach
is again the top-performer in terms of ISE, possibly due to the fact that the true counterfactual distributions are mixtures of normals. Table~\ref{t:sim2to4} displays the RMSE and AAB of QTEs for all approaches. The results show that the proposed approach achieves similar performance as DPM-BART in QTE estimation. Compared to SPQR-BART and SPQR-X, the proposed approach reduces the bias of almost all estimated QTEs, giving clear evidence of the advantage of double adjustment. The overall lower bias of the SPQR-DS and DPM-BART estimators compared to the TS estimator also suggests that the information of the treatment assignment mechanism is better utilized through PS adjustment than IPW in this specific setting.

\begin{figure}[tb]
\centerline{\includegraphics[width=0.8\linewidth]{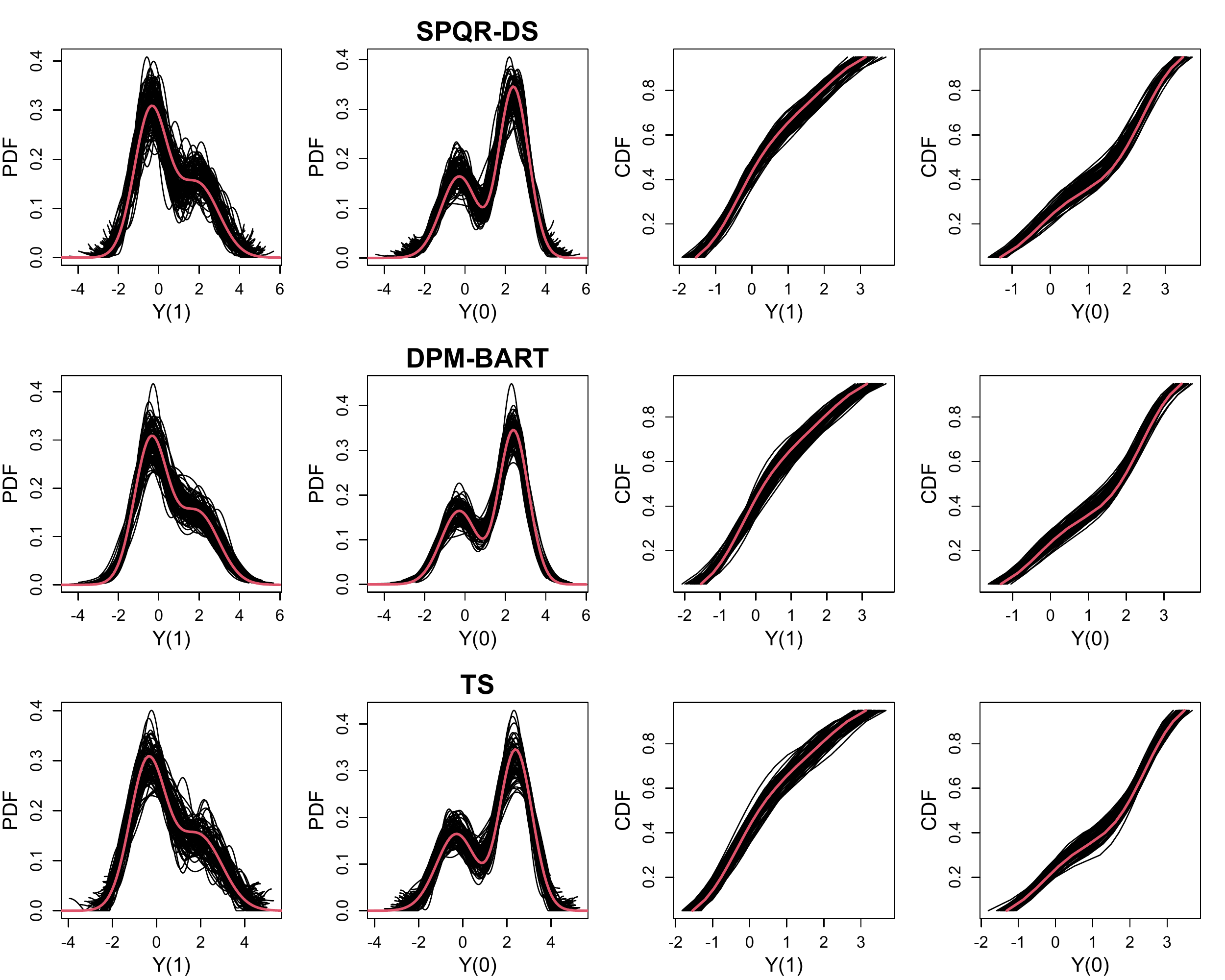}}
\caption{Simulation 2: Estimated (black lines) and true (red line) distributions of potential outcomes for 100 replicates.}\label{f:ex2}
\end{figure}

\begin{table}[!tb]
    \small
    \centering
    \caption{Simulation 2 -- 4. AAB and $\tau$-specific RMSE of QTE for all approaches. AAB is calculated using all 19 quantiles, and standard deviation is given in parentheses.}
    \label{t:sim2to4}
    \midsepremove
    \begin{tabular}{@{}*{8}c@{}}\toprule
     & & \multicolumn{6}{c}{RMSE of $\widehat{\Delta}_{\text{QTE}}(\tau)$}\\ 
    \cmidrule(lr){3-8}
     & $\tau$ & SPQR-DS  & SPQR-BART & SPQR-X & DPM-BART  & DPM-X & TS\\
    \midrule

    \multicolumn{3}{l}{Simulation 2}\\
                       &  0.1   & 0.14 & 0.13 & 0.15 & 0.15 & -- & 0.16 \\
                       &  0.25   & 0.16 & 0.16 & 0.19 & 0.17 & -- & 0.17 \\
                       &  0.5   & 0.20 & 0.23 & 0.23 & 0.21 & -- & 0.2 \\
                       &  0.75   & 0.21 & 0.26 & 0.24 & 0.20 & -- & 0.21 \\
                       &  0.9   & 0.18 & 0.20 & 0.19 & 0.19 & -- & 0.23 \\
    \cmidrule(lr){2-8}                  
    & AAB &  \textbf{0.15 (0.07)} & 0.16 (0.09) & 0.17 (0.09) & \textbf{0.15 (0.07)} & -- & 0.16 (0.07) \\
    \multicolumn{3}{l}{Simulation 3}\\
                       &  0.1   & 0.08 & 0.08 & 0.08 & 0.08 & 0.08 & 0.10 \\
                       &  0.25   & 0.09 & 0.12 & 0.09 & 0.10 & 0.11 & 0.11 \\
                       &  0.5   & 0.12 & 0.17 & 0.12 & 0.16 & 0.17 & 0.17 \\
                       &  0.75   & 0.16 & 0.19 & 0.17 & 0.17 & 0.16 & 0.21 \\
                       &  0.9   & 0.13 & 0.14 & 0.13 & 0.13 & 0.18 & 0.13 \\
    \cmidrule(lr){2-8}                  
    & AAB &  \textbf{0.09 (0.04)} & 0.12 (0.05) & 0.10 (0.04) & 0.11 (0.04) & 0.12 (0.04) & 0.12 (0.05) \\
    \multicolumn{4}{l}{Simulation 4}\\
                       &  0.1   & 0.01 & 0.01 & 0.01 & 0.01 & 0.02 & 0.01 \\
                       &  0.25   & 0.02 & 0.02 & 0.02 & 0.03 & 0.03 & 0.02 \\
                       &  0.5   & 0.03 & 0.03 & 0.04 & 0.04 & 0.06 & 0.03 \\
                       &  0.75   & 0.06 & 0.06 & 0.06 & 0.06 & 0.06 & 0.06 \\
                       &  0.9   & 0.12 & 0.11 & 0.12 & 0.11 & 0.13 & 0.17 \\ 
    \cmidrule(lr){2-8}                  
    & AAB &  \textbf{0.04 (0.02)} & \textbf{0.04 (0.02)} & \textbf{0.04 (0.02)} & \textbf{0.04 (0.02)} & 0.05 (0.02) & \textbf{0.04 (0.02)}\\ \bottomrule
    \end{tabular}
    \begin{tablenotes}
        \item[a] Note: best-performing models with the smallest average AAB are in bold fonts.
    \end{tablenotes}
    \midsepdefault
\end{table}

\subsection{Simulation 3}
Each subject is associated with 4 continuous confounders. The true PS model is a NN model with one hidden layer and five hidden neurons. One potential outcome follows a normal distribution and the other follows a skew normal distribution. The exact form of the true model is given in Appendix B.

Figure~\ref{f:ex3} displays the true and
estimated counterfactual PDFs and CDFs from the
100 replicates for SPQR-DS, DPM-BART and TS. The results show that all three approaches correctly estimate the counterfactual distributions. Figure~\ref{f:ex_ise}(d) displays the boxplots of ISE for all approaches. The proposed approach performs the best in estimating the control density, whereas the DPM-BART approaches performs the best in estimating the treatment density. The results show that even though the outcome-PS dependence relationship is quite complex, the BART probit model is flexible enough to approximate it reasonably well. Table 2 displays the RMSE and
AAB of QTEs for all approaches. The results show that the proposed approach achieves
the best performance in QTE estimation. Comparison with SPQR-BART and DPM-BART shows that the proposed approach benefits from additionally adjusting for the individual covariates which help explain the variability of outcome residual after adjusting for the PS. Comparison with SPQR-X and DPM-X shows that the inclusion of the PS improves the estimation of the conditional distribution in presence of strong confounding.

\begin{figure}[tb]
\centerline{\includegraphics[width=0.8\linewidth]{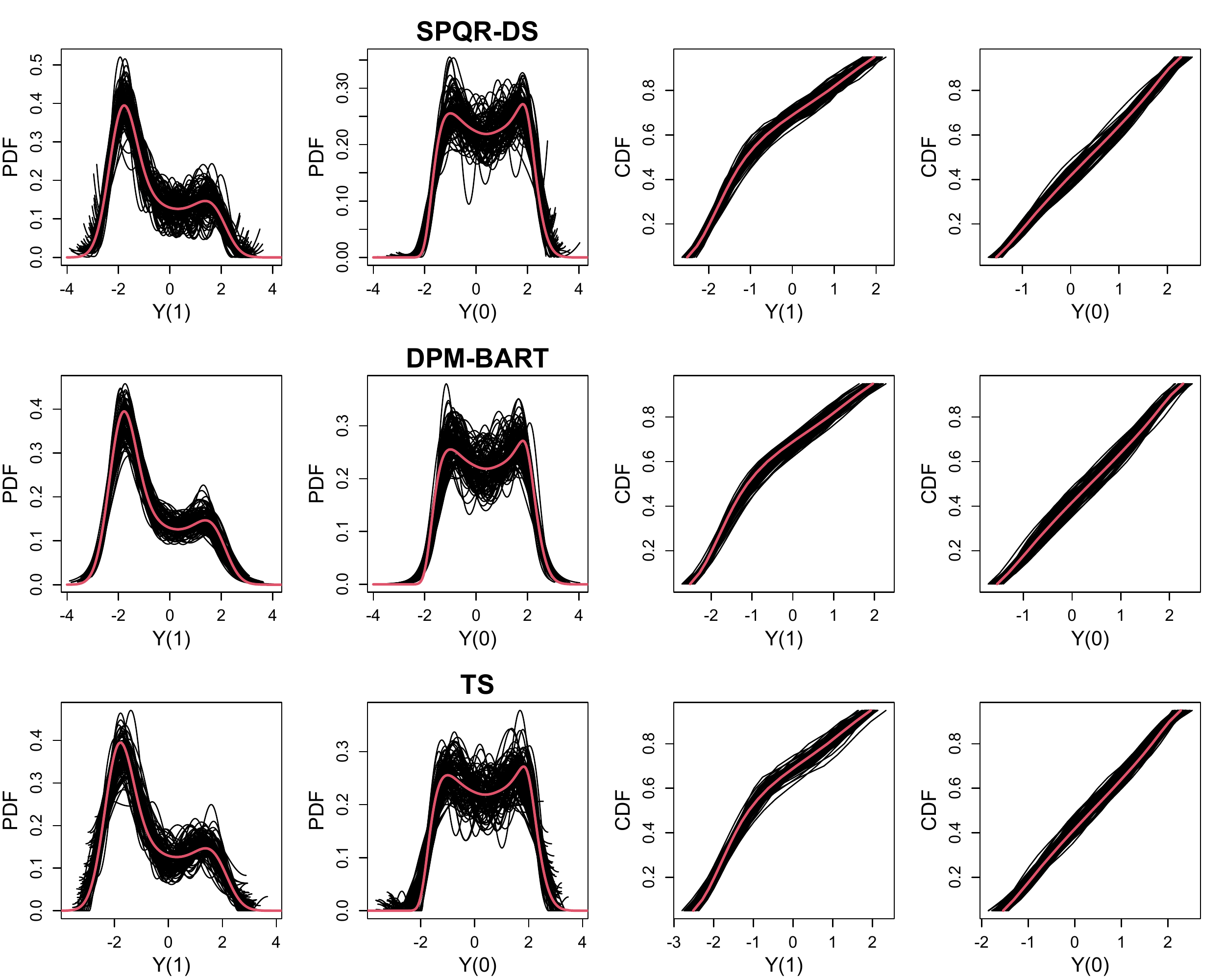}}
\caption{Simulation 3: Estimated (black lines) and true (red line) distributions of potential outcomes for 100 replicates.}\label{f:ex3}
\end{figure}

\subsection{Simulation 4}
Given that the true counterfactual distributions in Simulation 1 to 3 are all based on normal distributions or skew normal distributions, it is not surprising to see that the DPM-BART approach consistently yields excellent performance in estimating the counterfactual distributions. We show in this simulation example that the use of a parametric mixing distribution can become restrictive in certain cases, whereas the proposed approach is more flexible with its use of spline basis.

For simplicity, we set the true $\pi(\bm{X})$ to be $0.5$ so that the data represent a randomized experiment setup. Each subject is associated with 5 continuous covariates that have no effect on the outcome distribution. The potential outcomes have exponential distributions with different rate parameters. The exact form of the true model is
\begin{equation*}
    \begin{split}
    X_j&\sim\ \mathcal{U}(-2,2)\hspace{5mm} j=1,...,5\\
    T|\bm{X}&\sim \mathcal{B}ern(0.5)\\
    Y(0)|\bm{X}&\sim \mathcal{E}xp(2)\\\
    Y(1)|\bm{X}&\sim \mathcal{E}xp(4).
    \end{split}
\end{equation*}
The exponential distribution is difficult to approximate with a normal mixture since it does not have a well defined mode.

\begin{figure}[tb]
\centerline{\includegraphics[width=0.8\linewidth]{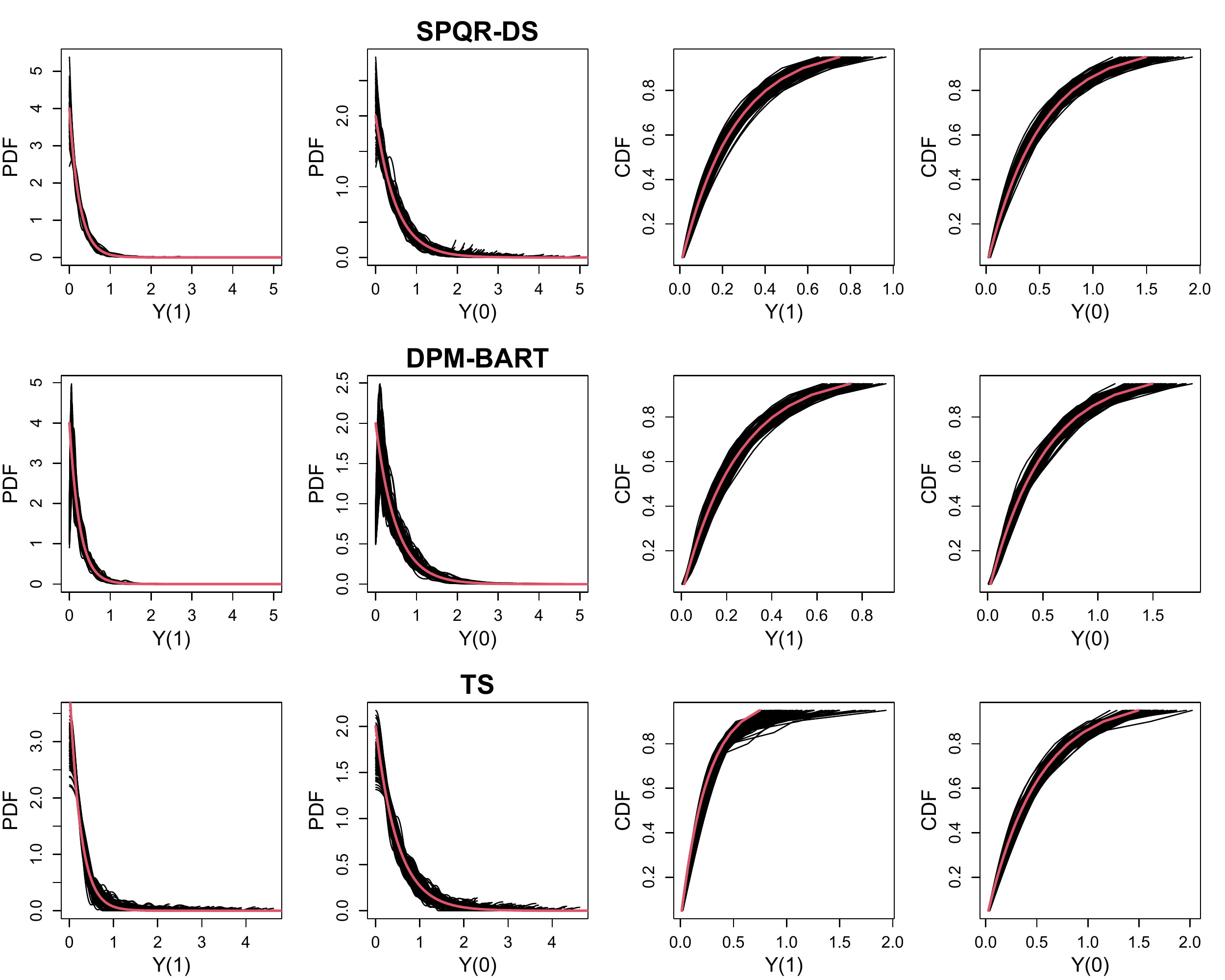}}
\caption{Simulation 4: Estimated (black lines) and true (red line) distributions of potential outcomes for 100 replicates.}\label{f:ex4}
\end{figure}

Figure~\ref{f:ex4} displays the true and estimated counterfactual PDFs and CDFs from the 100 replicates for
SPQR-DS, DPM-BART and TS. The results show that the proposed approach and the TS approach correctly capture the monotonic nature of the underlying densities, whereas the DPM-BART gives a misleading description of the densities as it tries to fit a unimodal distribution to the data. However, the DPM-BART estimates the CDFs correctly. This shows that the counterfactual PDFs indeed provide more nuanced information than the CDFs and are more difficult to estimate accurately. Thus, a model that can accurately estimate both the counterfactual PDFs and CDFs is crucial for correct inference on the counterfactual distributions. Figure~\ref{f:ex_ise}(e) displays the boxplots of ISE for all approaches. The results show that the SPQR-based approaches yield smallest ISE, whereas the DPM-based approaches yield the largest ISE. Table 2 displays the RMSE and AAB of QTEs for all approaches. The results show that all approaches yield similar overall performance on QTE estimation. However, compared to the other approaches, the TS approach yields significantly worse performance on estimating upper-tail QTEs. This again shows that the TS approach has poor boundary properties due to the oscillating nature of cosine series.

\section{Data application}
\label{s:data}
In this section, we apply our proposed method to estimate the causal effect of maternal smoking on birth weight distribution. In the literature, many studies have shown that low birth weight is associated with increased risk of health problems after birth and long-term economic cost due to medication, and that maternal smoking is one of the major modifiable risk factor for low birth weight \citep{Abrevaya2001,ngwira2015}. Consequently, there has been a great deal of interest in studying the causal effect of maternal smoking on infant birth weight \citep[e.g.,][]{abrevaya2015,huang2020,xie2020,zhou2021}. We adopt a data set  based on records between 1988 and 2002 by the North Carolina
Center Health Services. This data set was analyzed by \citet{abrevaya2015} in the context of CATE estimation and can be
downloaded from Prof. Leili’s website (\url{http://www.personal.ceu.hu/staff/Robert_Lieli/cate-birthdata.zip}). We focus on White and first-time mothers, and form a random sub-sample with sample size $n=5000$. The outcome $Y$ is birth weight measured in grams and the treatment $T$ is a binary indicator of maternal smoking (1: Yes, 0: No). We are interested in estimating the distribution of birth weight had all versus none of the mothers smoked in the entire population, as well as the treatment effect of maternal smoking on different birth weight quantiles. To ensure \textbf{SITA} holds, we choose a large set of variables as $\bm{X}$, including mother's age, education level (in years), the number of prenatal visits, the number of prenatal visits within the first trimester; and indicators for baby's gender, mother's marital status, gestational diabetes, hypertension, amniocentesis, ultrasound exams and alcohol use.

\begin{figure}[tb]
\centerline{\includegraphics[width=0.8\linewidth]{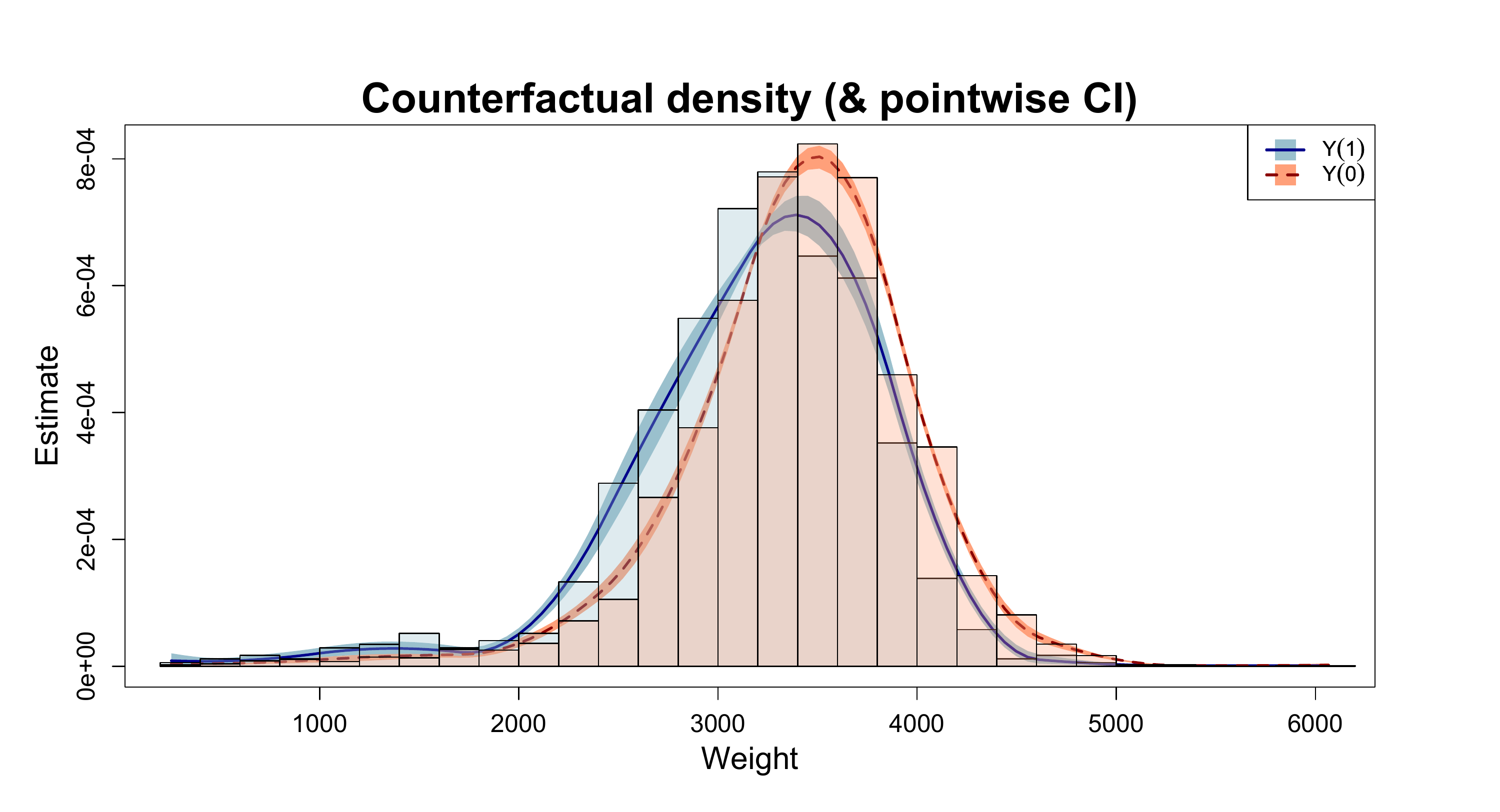}}
\caption{Estimated counterfactual birth weight density for infants born to smoking $(T=1)$ and nonsmoking $(T=0)$ mothers.}\label{f:bw_den}
\end{figure}

The response variable and all continuous covariates are mapped to the unit interval
using min-max normalization. We use five posterior samples from BART probit as an informative prior for the PS, and fit the proposed model based on NN with one or two hidden layers to the data. We run the MCMC for 10000 iterations and save every 10th iteration after discarding 1000 iterations as burn-in. The number of basis functions and hidden neurons are selected from $K=\{8,10,12\}$, $V_1=\{5,8,10,20\}$, $V_2=\{5,8,10,20\}$ using WAIC. Figure~\ref{f:bw_den} shows the estimated counterfactual birth weight densities using $K=8$ and $V_1=10$, along with $95\%$ pointwise CIs, overlaid on the histogram of the observed data. The plot shows that the estimated densities fit the data well. Compared to the birth weight density of nonsmoking mothers, the birth weight density of smoking mothers is more left-skewed with significant higher density in the lower-tail range ($< 3000$ grams), indicating that smoking mothers have a higher probability of giving birth to infants with low birth weight.

\begin{figure}[tb]
\centerline{\includegraphics[width=0.8\linewidth]{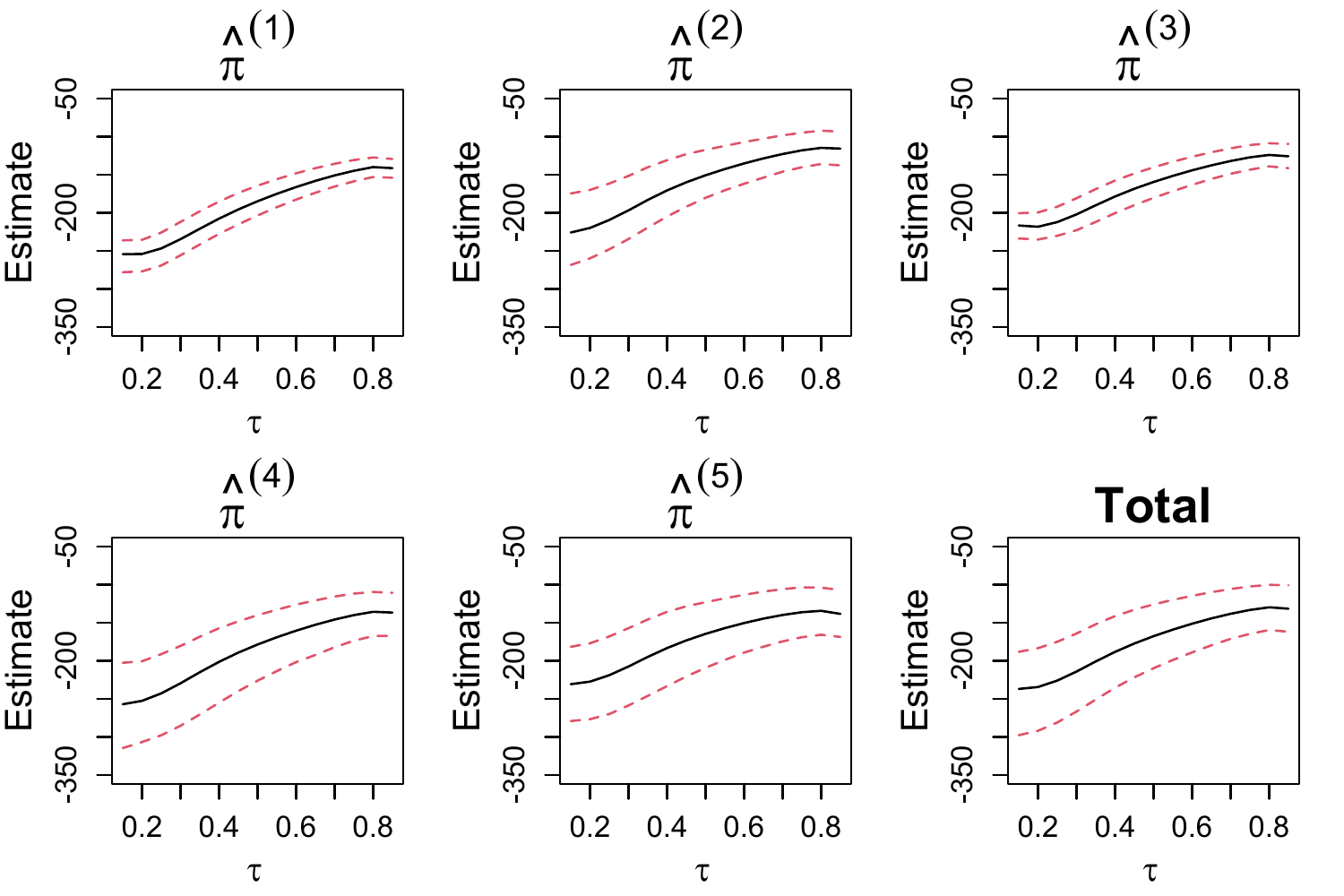}}
\caption{Treatment effect of maternal smoking on birth weight quantiles.}\label{f:bw_qte}
\end{figure}

To more precisely understand how maternal smoking affect the birth weight distribution, we estimate its QTE at quantile levels $\tau=0.15,0.20,...,0.85$. The QTEs estimated using posterior samples associated with each of the five estimated PS, as well as the QTE estimated using all posterior samples, are plotted in Figure~\ref{f:bw_qte}. The plots show that the estimated QTEs are consistent across different estimates of the PS, indicating convergence of the proposed approach. The estimated QTE seems to be monotonically increasing across the quantile domain, indicating that maternal smoking has the most significant effect on the lower tail of the distribution, which is the most critical aspect of the distribution. The $15$th QTE of maternal smoking is $-236$ gram (95\% CI: $-188.2$, $-297.8$), suggesting that maternal smoking leads to a $236$-gram decrease in the $15$th quantile of the birth weight distribution.

\section{Conclusion}
\label{sec:conc}

In this paper, we proposed a Bayesian semiparametric model that allows simultaneous estimation of any functionals of the counterfactual distributions in presence of confounding. Differ from existing works that consider regression adjustment for either the scalar PS or the full-vector of covariates, we proposed to adjust for a double-balancing score that augments PS adjustment with individual covariates to fully utilize the observed information. We then outlined a Bayesian inference framework that incorporates uncertainty of estimating the PS, the outcome distribution conditional on the balancing score, and the covariate distribution. Carefully designed simulations provides empirical evidence that the use of double-balancing score for confounding adjustment improves performance over adjusting for any individual score alone. First, the additional PS adjustment can reduce any residual confounding if the outcome-covariate relationship is complex and not fitted well. Second, the covariate adjustment can further control the variability of the outcome residual after adjusting for the PS, rendering the estimator more efficient. In regards to counterfactual distribution estimation, the proposed approach correctly estimates PDFs and CDFs of various shapes and achieves comparable and sometimes even better performance than the state-of-art DPM of normals. Furthermore, the use of SPQR as the conditional distribution estimator allows a broader model coverage than DPM of normals and better boundary properties than truncated series estimator. In regards to QTE estimation, the proposed model consistently yields the lowest bias among all competing models.

We applied the proposed approach to estimate the causal effect of maternal smoking on infants' birth weight. The estimated counterfactual densities show that smoking leads to substantial left-skewness of the counterfactual distribution and higher probability of giving birth to infants with low birth weight. The estimated QTEs suggest that the effect of smoking is most significant on the lower-tail quantiles of the birth weight distribution, further increasing the probability of having underweight infants. The results are consistent with the findings from other studies that maternal smoking might leads to lower birth weight \citep{abrevaya2015,huang2020,xie2020}.

Since the proposed approach always includes the full-vector of covariates, it ultimately will suffer from curse of dimensionality. A possible improvement could be to apply sufficient dimension reduction \citep[SDR;][]{ma2013} to the covariates and construct the double-balancing score by augmenting the PS with the projected covariates. However, incorporating the uncertainty of SDR can be computationally challenging. Another possible extension is to come up with a joint estimation framework for the PS and the outcome distribution. As mentioned in Section~\ref{s:prior}, we chose a sequential approach purely for the sake of convenience. This is fundamentally different from the reason behind the work of \citet{xu2018} where the PS has to be estimated separately to reduce bias arising from feedback issue \citep{zigler2013}. 

\section*{Funding}
This work was supported by the the National Science Foundation (DMS2152887 to B.J.R, DMS1811245 to S.Y.) and by the National Institutes of Health (R01ES031651-01 to B.J.R, 1R01AG066883 to S.Y.).

\section*{Appendices}
\section*{Appendix A. Detail of Simulation 2}
\label{a:A}

The exact form of the true model is
\begin{equation*}
    \begin{split}
        X_j&\sim\begin{cases}
        \mathcal{U}(0,1)\hspace{28pt} j=1,2,3\\
        \mathcal{U}(1,2)\hspace{28pt} j=4,5,6\\
        \mathcal{B}ern(0.5)\ \ \ j=7,...,12
        \end{cases}\\
        T|\bm{X}&\sim\mathcal{B}ern\lpth Z\rpth\\
        Y(0)|\bm{X}&\sim\sqrt{Z}\mathcal{N}\lpth2Z^2+X_4+X_3,0.5^2\rpth+\lpth1-\sqrt{Z}\rpth\mathcal{N}\lpth Z^2+X_2-\sum_{j=1}^3X_j^2,0.8^2\rpth\\
        Y(1)|\bm{X}&\sim0.6\mathcal{N}(-Z,0.8^2)+0.4\mathcal{N}(X_5+Z,1)
    \end{split}
\end{equation*}
where $Z=\text{expit}\lpth-2.125+0.5X_1X_4+X_2X_5+\sum_{j=1}^6X_jX_{j+6}\rpth$.

\section*{Appendix B. Detail of Simulation 3}
\label{a:B}
The exact form of the true model is
\begin{equation*}
    \begin{split}
        \begin{pmatrix}
        X_1\\
        X_2\\
        X_3\\
        X_4
        \end{pmatrix}&\sim\mathcal{N}\lsbk
        \begin{pmatrix}
        0\\
        0\\
        0\\
        0
        \end{pmatrix},
        \begin{pmatrix}
        1&0.5&0.2&0.3\\
        0.5&1&0.7&0\\
        0.2&0.7&1&0\\
        0.3&0&0&1
        \end{pmatrix}\rsbk\\
        T|\bm{X}&\sim\mathcal{B}ern\lpth\text{expit}\lcbk\bm{W}^o\times\text{tanh}\lpth\bm{W}^h\bm{X}^\top+\bm{b}^h\rpth^\top+b^o\rcbk\rpth\\
        Y(0)|\bm{X}&\sim\mathcal{SN}\lpth2\text{tanh}\lpth X_2-X_3+0.5X_4\rpth,0.5,3\rpth\\
        Y(1)|\bm{X}&\sim\mathcal{N}\lpth2\text{tanh}\lpth X_1+0.5X_2-X_3^2\rpth,0.5^2\rpth
    \end{split}
\end{equation*}
where
\begin{equation*}
\begin{split}
    \bm{W}^h&=\begin{pmatrix}
    -0.99&-1.1&-0.14&-0.26\\
    -0.18&0.03&-1.45&-0.07\\
    -0.44&0.19&0.86&0.36\\
    -1.07&0.67&-0.58&-0.13\\
    0.12&-0.37&0.47&1.25
    \end{pmatrix},\ \ 
    \bm{b}^h=\begin{pmatrix}
    0.96\\
    0.64\\
    0.74\\
    -0.46\\
    0.21
    \end{pmatrix}\\
    \bm{W}^o&=\begin{pmatrix}
    -0.15\\
    0.3\\
    -0.004\\
    -0.21\\
    -0.88
    \end{pmatrix},\ \ b^o=-0.05.
\end{split}
\end{equation*}

\bigskip
\begin{center}
{\large\bf SUPPLEMENTARY MATERIAL}
\end{center}





\bibliographystyle{jabes}

\bibliography{BSPQTE}

\end{document}


\def\spacingset#1{\renewcommand{\baselinestretch}%
{#1}\small\normalsize} \spacingset{1}


\if0\blind
{
  \title{\bf Appendix of ``A Bayesian Semiparametric Method For Estimating Causal Quantile Effects"}
  \author{Steven G. Xu, Shu Yang and Brian J. Reich\\
    Department of Statistics, North Carolina State University}
  \maketitle
} \fi

\if1\blind
{
  \bigskip
  \bigskip
  \bigskip
  \begin{center}
    {\LARGE\bf Title}
\end{center}
  \medskip
} \fi

\bigskip

\section*{Web Appendix A}
\label{a:A}

\renewcommand{\thesection}{\Alph{section}}
\setcounter{section}{1}
This section includes simulation results that are not presented in the main paper due to space limitations. Figures~S.1 -- S.4 display the estimated and true counterfactual PDFs and CDFs for the SPQR-DS approach, the DPM-BART approach, and the TS approach. Figure~S.5 -- S.8 display the boxplots of integrated squared error (ISE) of counterfactual density estimation for all approaches.

\renewcommand{\thefigure}{S.1}
\begin{figure}[tbh]
\centerline{\includegraphics[width=\linewidth]{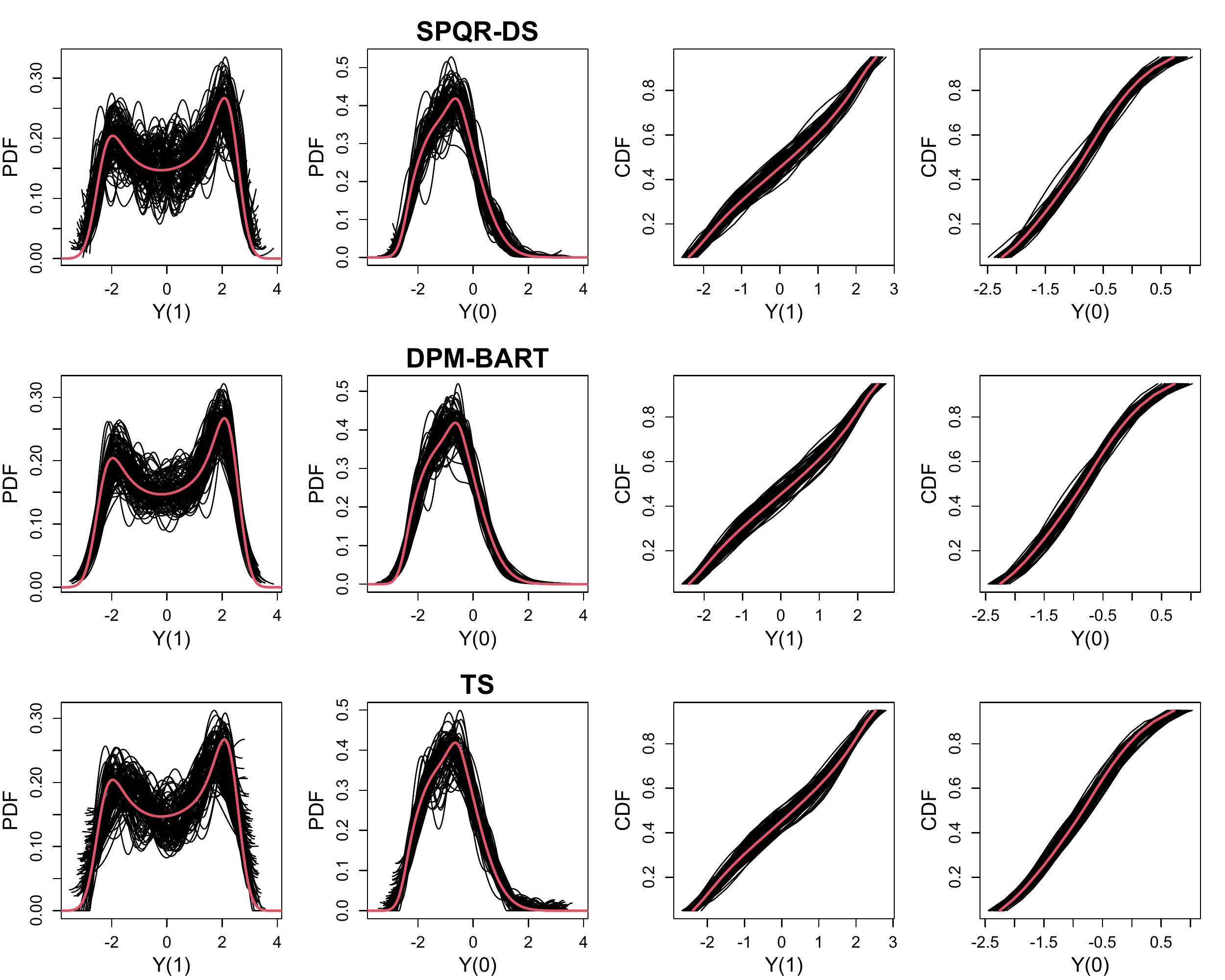}}
\caption{Simulation Example 1, $J=0$: Estimated PDFs and CDFs (black lines) of potential outcomes compared to the ground truth (red line), for 100 replicates for the SPQR-DS approach, the DPM-BART approach, and the TS approach.}\label{f:ex1_0}
\end{figure}

\renewcommand{\thefigure}{S.2}
\begin{figure}[tbh]
\centerline{\includegraphics[width=\linewidth]{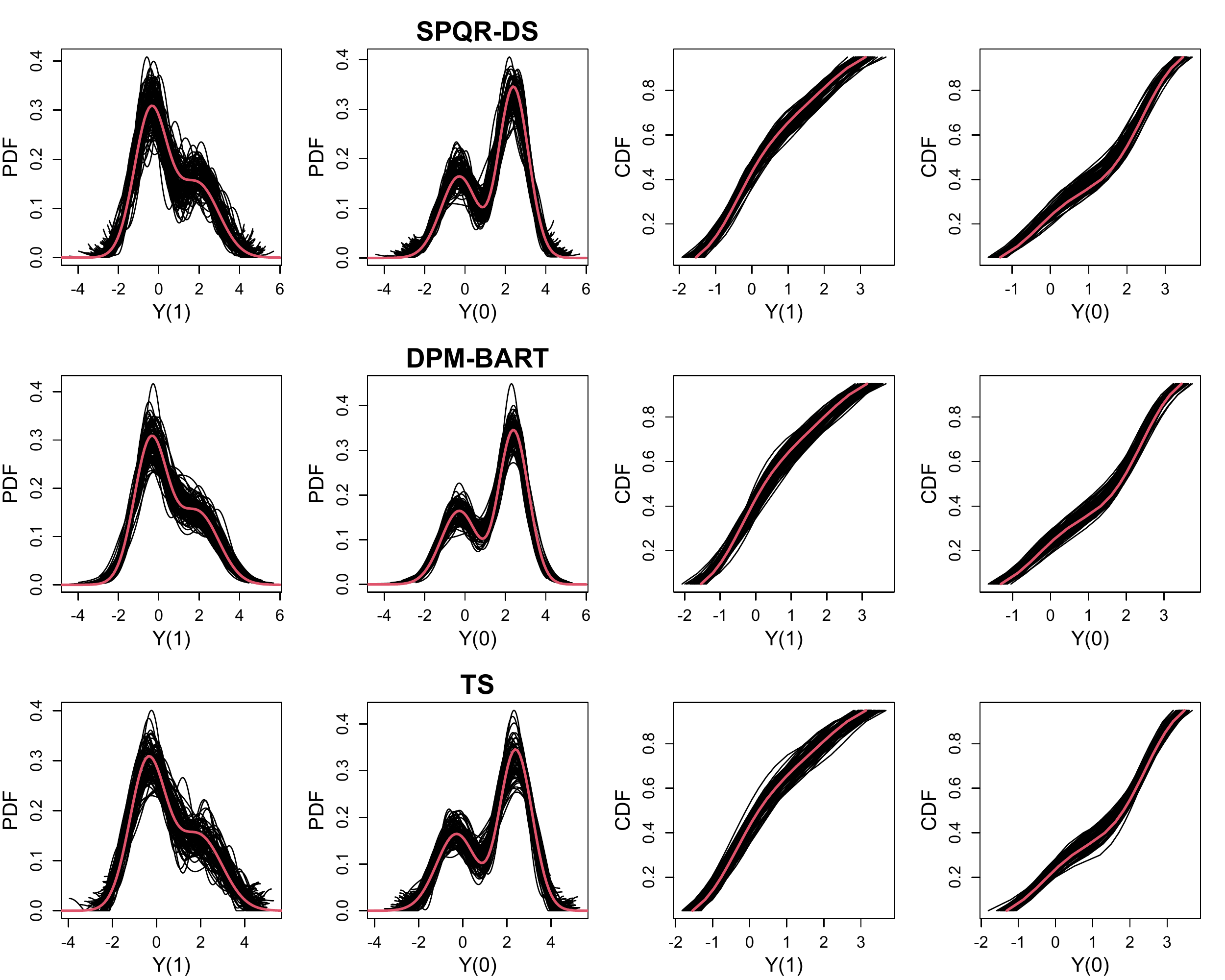}}
\caption{Simulation Example 2: Estimated PDFs and CDFs (black lines) of potential outcomes compared to the ground truth (red line), for 100 replicates for the SPQR-DS approach, the DPM-BART approach, and the TS approach.}\label{f:ex2}
\end{figure}

\renewcommand{\thefigure}{S.3}
\begin{figure}[tbh]
\centerline{\includegraphics[width=\linewidth]{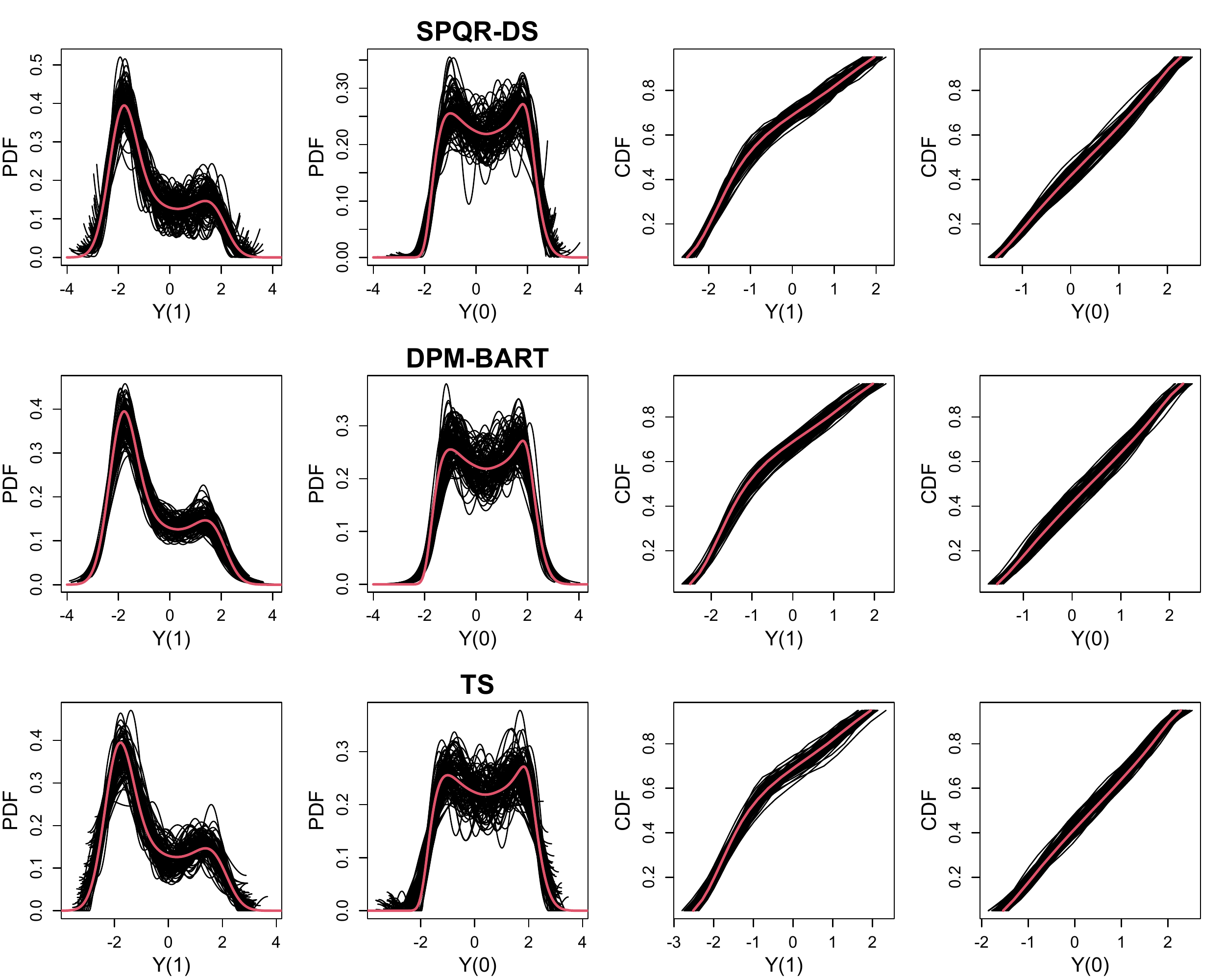}}
\caption{Simulation Example 3: Estimated PDFs and CDFs (black lines) of potential outcomes compared to the ground truth (red line), for 100 replicates for the SPQR-DS approach, the DPM-BART approach, and the TS approach.}\label{f:ex3}
\end{figure}

\renewcommand{\thefigure}{S.4}
\begin{figure}[tbh]
\centerline{\includegraphics[width=\linewidth]{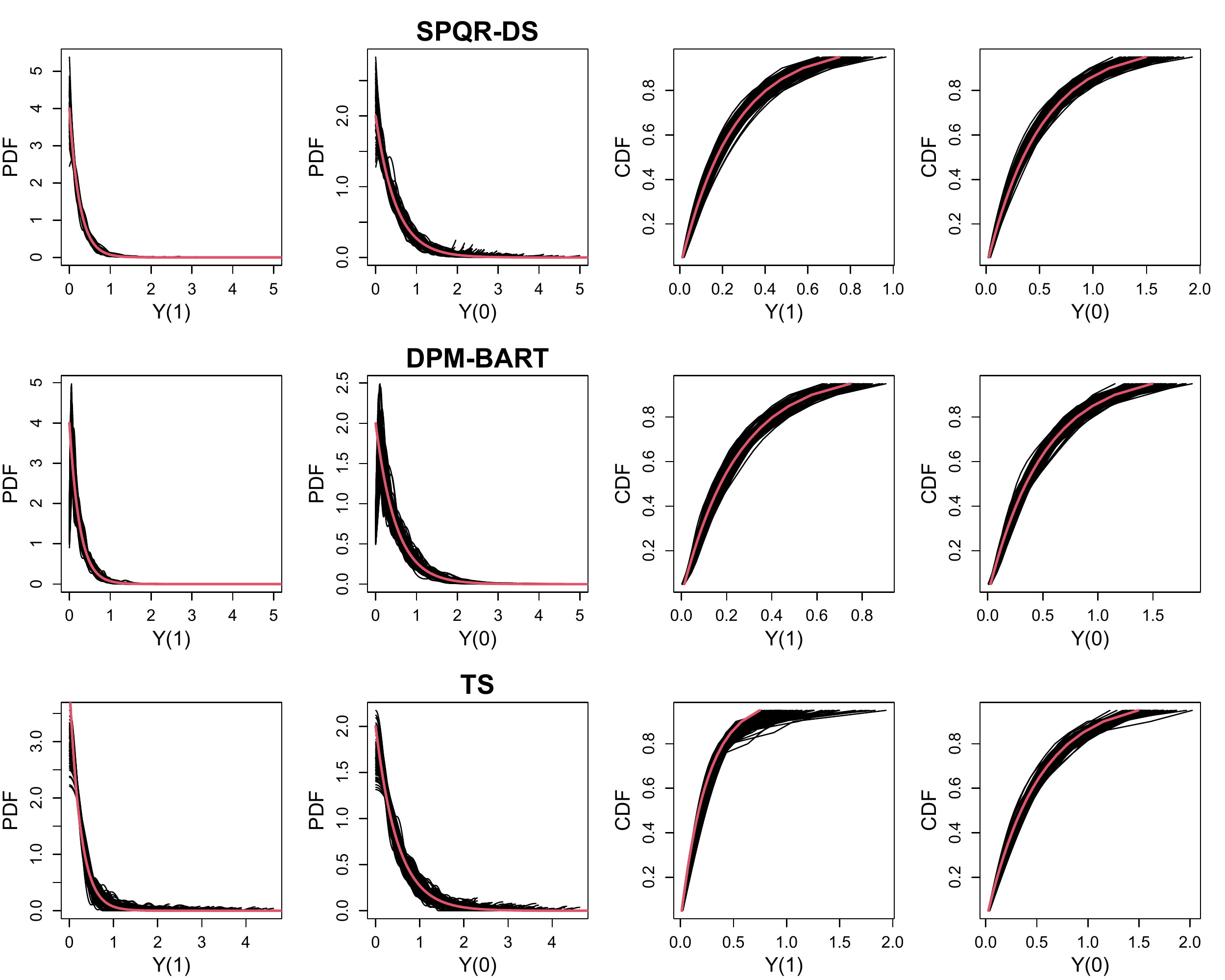}}
\caption{Simulation Example 4: Estimated PDFs and CDFs (black lines) of potential outcomes compared to the ground truth (red line), for 100 replicates for the SPQR-DS approach, the DPM-BART approach, and the TS approach.}\label{f:ex4}
\end{figure}

\renewcommand{\thefigure}{S.5}
\begin{figure}[tbh]
\centerline{\includegraphics[width=\linewidth]{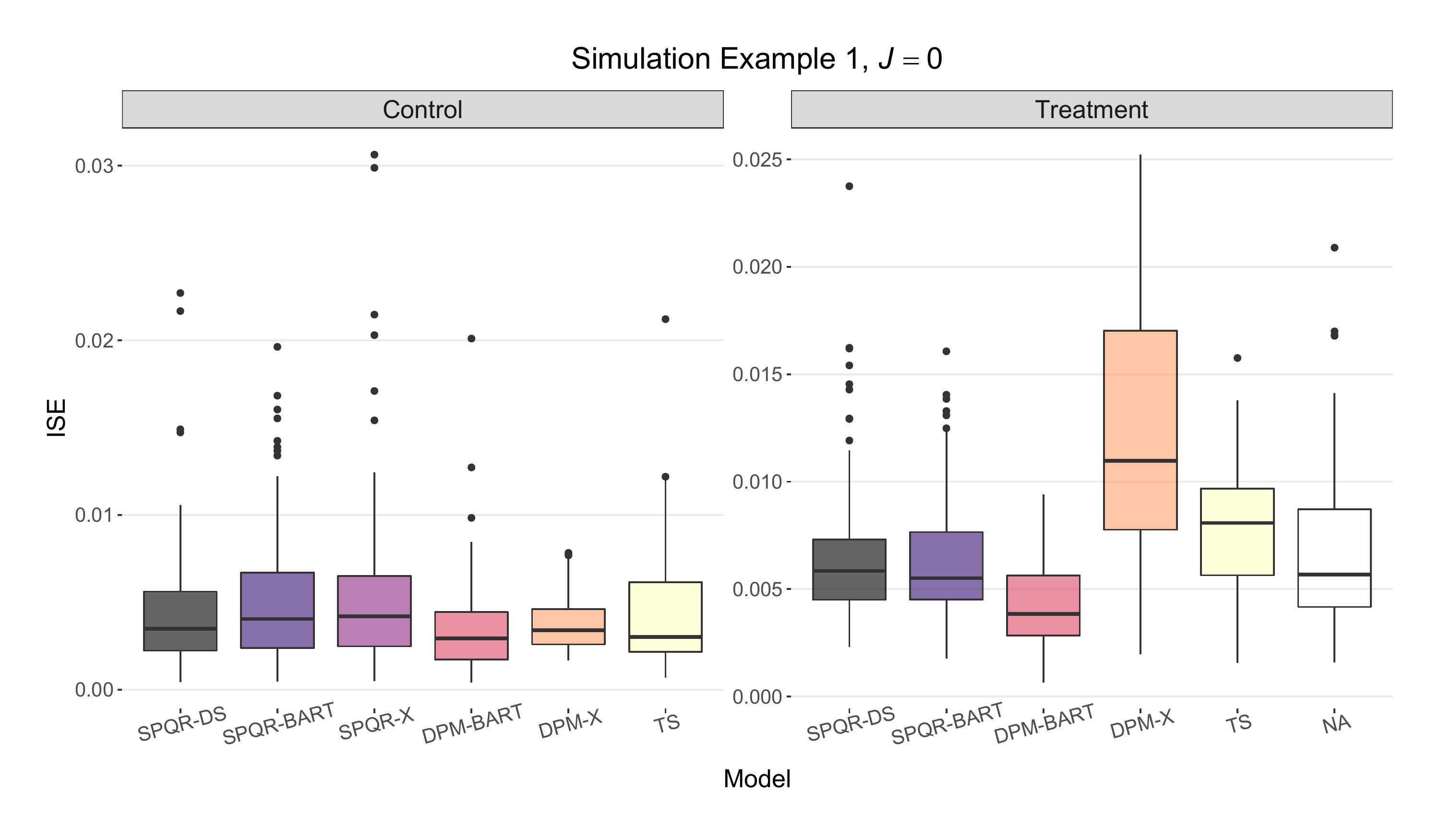}}
\caption{Simulation Example 1, $J=0$: ISE of estimated counterfactual densities for 100 replicates for all approaches.}\label{f:ex1_0_ise}
\end{figure}

\renewcommand{\thefigure}{S.6}
\begin{figure}[tbh]
\centerline{\includegraphics[width=\linewidth]{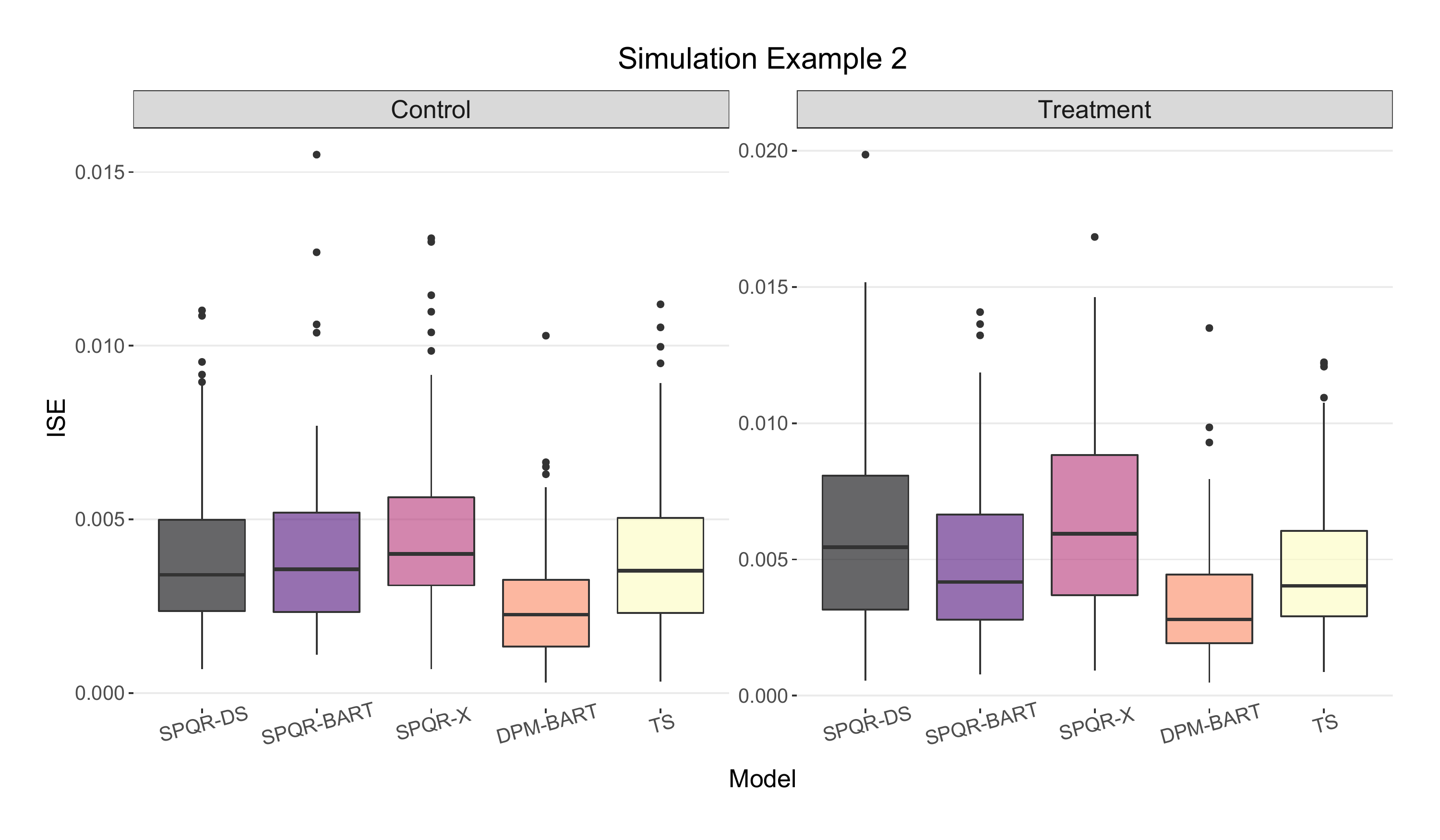}}
\caption{Simulation Example 2: ISE of estimated counterfactual densities for 100 replicates for all approaches except for DPM.}\label{f:ex2_ise}
\end{figure}

\renewcommand{\thefigure}{S.7}
\begin{figure}[tbh]
\centerline{\includegraphics[width=\linewidth]{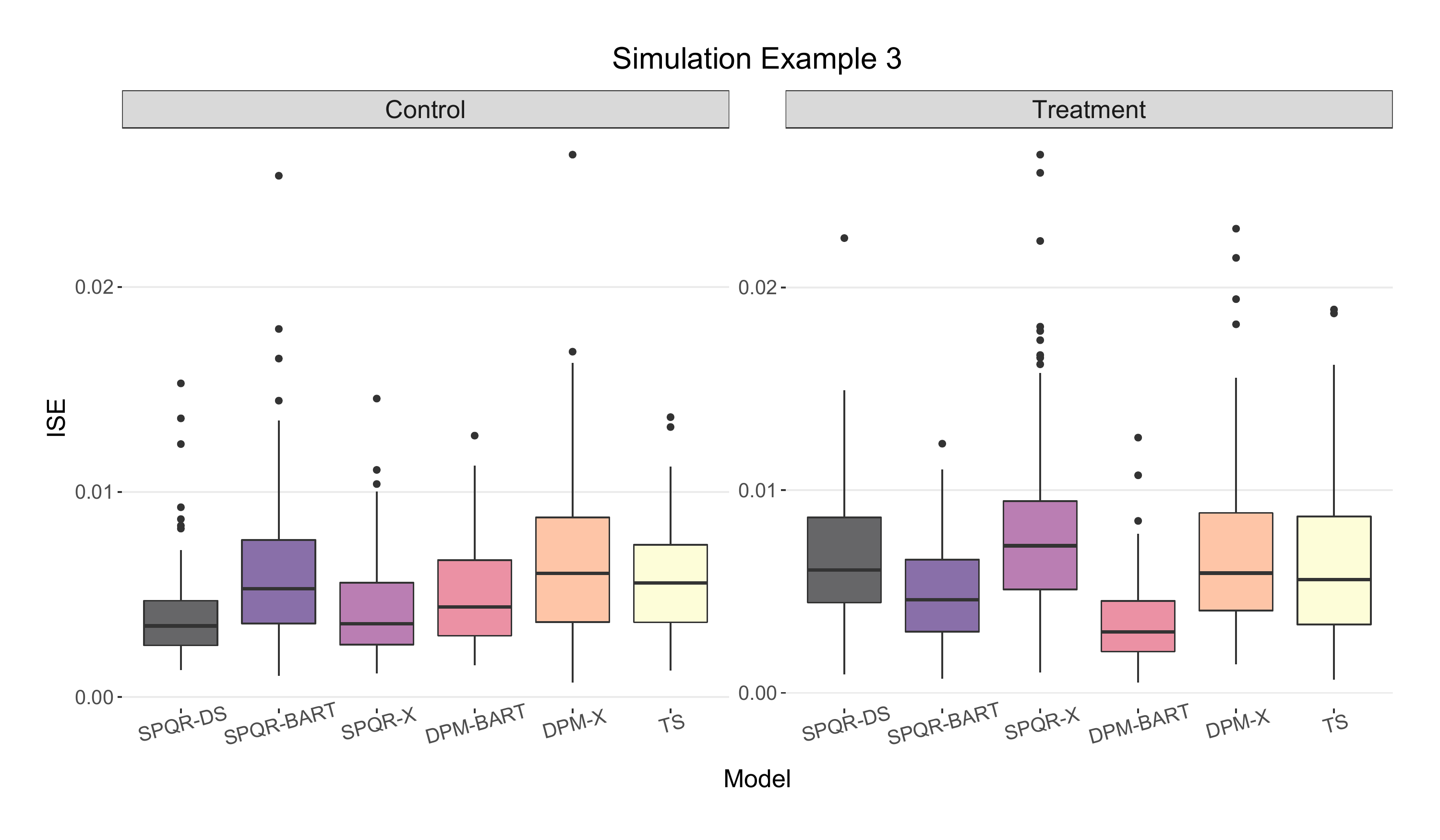}}
\caption{Simulation Example 3: ISE of estimated counterfactual densities for 100 replicates for all approaches.}\label{f:ex3_ise}
\end{figure}

\renewcommand{\thefigure}{S.8}
\begin{figure}[tbh]
\centerline{\includegraphics[width=\linewidth]{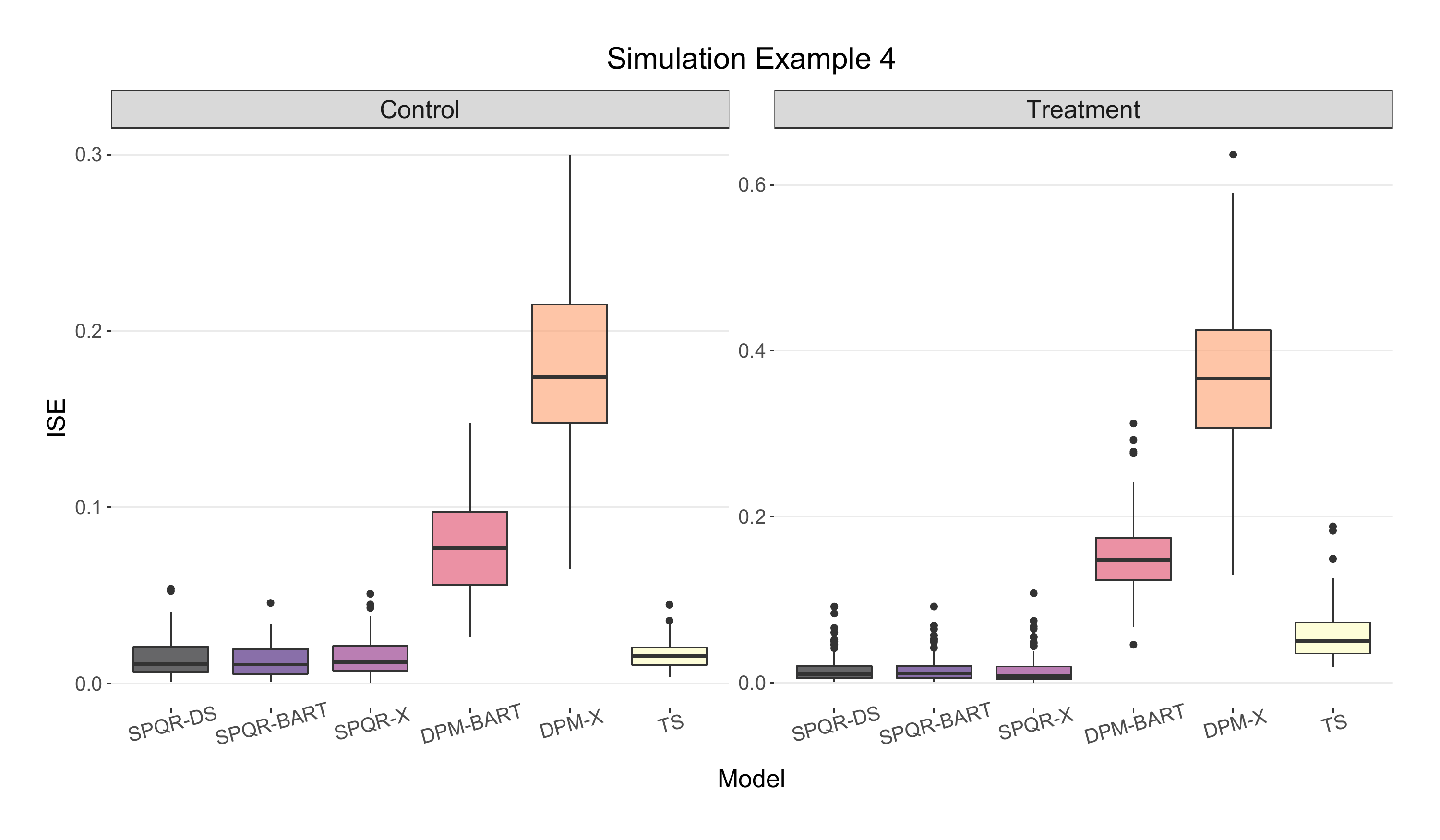}}
\caption{Simulation Example 4: ISE of estimated counterfactual densities for 100 replicates for all approaches.}\label{f:ex4_ise}
\end{figure}